\documentclass[onecolumn,aps,prd,preprintnumbers,showpacs,superscriptaddress,nofootinbib,amsmath,amssymb,floats,floatfix,showkeys,notitlepage]{revtex4-2}

\usepackage{orcidlink}
\usepackage{lipsum}
\usepackage{graphicx}
\usepackage{subfigure}
\usepackage{sans}
\usepackage{changes}
\usepackage{hyperref}
\hypersetup{colorlinks=true,linkcolor=blue,urlcolor=blue,citecolor=blue}
\usepackage[toc,page]{appendix}
\usepackage[normalem]{ulem}
\usepackage{adjustbox}
\usepackage{latexsym}
\usepackage{amsmath}
\usepackage{amssymb}
\usepackage{amsfonts}
\usepackage{dcolumn}
\usepackage{bm}
\usepackage{tikz}
\usepackage{bigints}
\usepackage{array,tabularx,multirow,booktabs}
\usepackage[tracking=true]{microtype}
\SetTracking{}{500}
\SetTracking{encoding={*}, shape=sc}{40}
\UseRawInputEncoding 
\allowdisplaybreaks

\usepackage{hyperref}
\usepackage{colortbl}
\usepackage{xcolor}

\begin{document} 

\title{Quasinormal modes and shadow in Einstein Maxwell power-Yang-Mills black hole 
}


\author{Angel Rincon
\orcidlink{0000-0001-8069-9162}
}
\email{angel.rincon@ua.es}
\affiliation{Departamento de F\'isica Aplicada, Universidad de Alicante, Campus de San Vicente del Raspeig, E-03690 Alicante, Spain}

\author{Gabriel G\'omez
\orcidlink{0000-0001-8069-9162}
}
\email{gabriel.gomez.d@usach.cl}
\affiliation{Departamento de F\'isica, Universidad de Santiago de Chile,Avenida V\'ictor Jara 3493, Estaci\'on Central, 9170124, Santiago, Chile}

\date{\today}
\begin{abstract}

In the present paper, we investigate the quasinormal modes of an Einstein-Maxwell power-Yang-Mills black hole in four dimensions, considering a specific value of the power parameter $p = 1/2$. This particular case represents a black hole with both Abelian and Non-Abelian charges and is asymptotically non-flat. We begin by deriving the effective potential for both a neutral massless particle and a  neutral Dirac particle using the aforementioned black hole solution. 
Subsequently, employing the sixth-order WKB approximation
method, we calculate the (scalar) quasinormal modes. Our numerical analysis indicates that these modes are stable within the considered parameter range. This result is also confirmed using the eikonal approximation.
Furthermore, we calculate the shadow radius for this class of BH and derive constraints on the electric and Yang-Mills charges ($Q, Q_{\rm YM}$) by using imaging observational data for Sgr A${^\star}$, provided by the Event Horizon Telescope Collaboration. We observe that as the electric charge $Q$ increases, the allowed range shifts towards negative values of $Q_{\rm YM}$. For instance, for the maximum value $Q\approx 1.1$ obtained, the allowed range becomes $-0.171 \lesssim Q_{\rm YM} \lesssim -0.087$ consistent with KECK and VLTI data, while still retaining a non-vanishing horizon.
\end{abstract}



\maketitle

\section{Introduction} \label{intr}

Black hole (BH) solutions play an essential role in classical and alternative theories of gravity, which seek to describe the properties of spacetime \cite{Misner:1973prb,Hartle:2003yu}. Any observational feature of BH can serve as a compelling tool for testing gravity theories, particularly at the event horizon scale, and thus contribute to establishing the true nature of gravity.  Interestingly, we have today convincing probes about the existence of BHs provided by the Event Horizon Telescope (EHT) and the Very Large Telescope global networks \cite{EventHorizonTelescope:2019dse,EventHorizonTelescope:2022wkp,EventHorizonTelescope:2019ths,EventHorizonTelescope:2022xqj}, the GRAVITY collaboration \cite{GRAVITY:2020gka}, and the LIGO-Virgo collaboration \cite{LIGOScientific:2017ync,LIGOScientific:2017vwq} among others observational evidences. What can we conclude about the nature of gravity from these results? First, the predictions of General Relativity (GR) are consistent with all observational data within the current uncertainties \cite{Will:2014kxa}. Second, theories beyond Einstein's theory can also explain the observed phenomena \cite{LIGOScientific:2020tif,Cardoso:2019rvt,EventHorizonTelescope:2020qrl,Vagnozzi:2022moj}. Hence, these results are encouraging for theories beyond GR, but no conclusive evidence has been found thus far to decisively support one theory over another.
\\
In order to get further insights into the nature of gravity, it is convenient to investigate the so-called \textit{quasinormal modes} (QNMs). Roughly speaking, QNMs  are energy dissipation modes of a perturbed BH. These modes characterize perturbations within a field that gradually diminish over time \cite{Berti:2009kk,Konoplya:2011qq}.
In simpler terms QNMs of a BH correspond to perturbed solutions of the field equations with complex frequencies, and their characteristics depend on the specific theoretical model under consideration. Consequently, a phenomenological strategy involves examining the properties and stability of QNMs associated with a particular BH solution. 
Notice that, from a theoretical perspective, perturbations within the spacetime of a BH can be examined through two distinct approaches. The initial approach involves introducing additional fields into the BH spacetime. Alternatively, the second method involves perturbing the underlying metric of the BH (referred to as the background). Furthermore, in the linear approximation, the first perturbation scenario can be simplified to the propagation of fields in the background of a BH.
In particular, for a given scalar field $\Phi$ with a given mass $\mu$ in the background of the metric $g_{\mu \nu}$, the master differential equation is the Klein-Gordon equation \cite{Konoplya:2011qq}. 
Solving such a differential equation analytically is generally quite challenging because the effective potential must be in a simplified form to obtain the corresponding solution. A few examples where the QNMs can be obtained analytically can be consulted at \cite{Leaver:1985ax,Musiri:2003ed,Aguayo:2022ydj,Panotopoulos:2018can,Siopsis:2008xz}. 
Therefore, in order to make progress, numerical/semi-analytical approaches should be considered. Although in the next section we will briefly mention some approaches to obtain the QNMs, we can highlight a few conventional methods for finding the corresponding solutions. For instance: i) the WKB semi-analytical approach, ii) the Mashhoon method, iii) the Chandrasekhar-Detweiler method, the Shooting method, among others.
So, even though the literature is vast, we can mention some recent works in which the QNMs, in GR and beyond, are calculated. For instance, the reader can consult \cite{Baruah:2023rhd,Becar:2022wcj,Fontana:2020syy,Konoplya:2023aph,Gogoi:2023fow,Lambiase:2023hng,Gogoi:2023kjt,Jawad:2023zlu,Balart:2023swp,Panotopoulos:2020mii,Rincon:2020iwy,Rincon:2020pne} and references therein.
Another intriguing feature of  BHs that has gained attention, particularly following the release of imaging observations of the Sgr A$^{\star}$ and M87$^{\star}$ BHs, is their shadow. The shadow refers to the dark region observed in the vicinity of a BH, surrounded by circular orbits of photons. This distinctive feature has been utilized as a potential discriminator to distinguish and differentiate between different BH solutions \cite{Vagnozzi:2022moj} (see also \cite{Uniyal:2022xnq}).
\\
In the framework of GR, BHs are characterized solely by three physical parameters: mass $M$, angular momentum $J$, and (electric) charge $Q$. This statement is known as the  "no-hair theorem" \cite{heusler_1996}. The more general case of BH solutions corresponds to the Kerr-Newman solution. However, BHs are commonly assumed t o be uncharged due, for instance, to charge neutralization process of astrophysical plasma. Nevertheless, recent observations provided by the EHT collaborations do not rule out the possibility that BH may carry some degree of charge \cite{EventHorizonTelescope:2021dqv,EventHorizonTelescope:2022xqj}, even in the context of more general theories of gravity \cite{Ghosh:2022kit}.
Building a viable BH solution beyond GR is a non-trivial task since the theory itself must prevent any pathological behavior. This includes preserving the hyperbolic character of the field equations, avoiding the propagation of unwanted perturbation modes, and addressing other issues that may arise at the theoretical level. This scientific program has been a highly active topic of research, driven by the possibility of detecting deviations from GR, which would provide valuable insights into the nature of gravity.
\\
We do not pretend to discuss here all classes of BH solutions. Instead, the main subject of this paper is on BH solutions involving non-Abelian gauge fields. The initial motivation for considering the coupling of the Yang-Mills theory to Einstein's gravity stems from the fact that they together provide a suitable framework for the existence of stationary, localized, and non-singular solutions known as solitons \cite{Bartnik:1988am} (for soliton solutions in a more general massive Yang-Mills theory, see, for example, Ref.~\cite{Martinez:2022wsy}). This is not achievable in separate scenarios. Subsequently, this idea was extended to construct BH solutions, resulting in BH with a Yang-Mills hair \cite{Volkov:1989fi,Volkov:1998cc}. Additional solutions involving the Yang-Mills theory can be found in \cite{Volkov:1998cc,Mazharimousavi:2008ap,Mazharimousavi:2009mb,Radu:2011ip,Herdeiro:2017oxy,Meng:2017srs,ElMoumni:2018bkl,HabibMazharimousavi:2008zz,Gomez:2023wei} and references therein. 
\\
Following the same principle employed in the study of nonlinear (Maxwell) electrodynamics \cite{Gurtug:2010dr,Xu:2014uka,Panotopoulos:2019tyg,Hendi:2010bk,Panotopoulos:2020zbj,Gonzalez:2021vwp,Rincon:2017goj,Panotopoulos:2017hns,Rincon:2018sgd,Rincon:2018dsq,Panotopoulos:2018rjx,Rincon:2021hjj,Rincon:2021gwd,Panotopoulos:2022bky,Hendi:2017uau,Panah:2022cay,Hendi:2017lgb,EslamPanah:2021xaf}, the Einstein-Yang-Mills BH solutions were further extended to include power Yang-Mills solutions characterized by a non-Abelian topological charge \cite{Mazharimousavi:2009mb} (see also \cite{ElMoumni:2018fml,Biswas:2022qyl,Stetsko:2020nxb} for further investigations). 
From a theoretical perspective, it is possible to couple the standard Maxwell theory and the power-law Yang-Mills theory to Einstein's gravity, thereby allowing for the existence of a more general class of BH with appealing features that can be potentially contrasted with observations. This proposal leads to BHs with both Abelian and non-Abelian charges or, equivalently, a modified version of the well-known Reissner-Nordstr\"{o}m BH solution. 
There are clear theoretical advantages to coupling a non-linear Yang-Mills field to gravity. First, it is possible to construct non-singular BH solutions \cite{HabibMazharimousavi:2008dm,Balakin:2015gpq}. Second, there exists a solution near a critical point in which the emergent BH is stable \cite{Mazharimousavi:2009mb,ElMoumni:2018fml,MasoumiJahromi:2023crl}, as demanded by thermodynamic arguments, unlike the canonical Yang-Mills field. Furthermore, the introduction of the non-linear Yang-Mills field in the Einstein framework is particularly useful for understanding thermodynamical aspects, such as phase space transitions, that emerge in the context of AdS/CFT \cite{ElMoumni:2018fml,Du:2022ckz}. This is due to the similarity between entanglement entropy and Bekenstein-Hawking entropy. This correspondence provides a powerful tool to study quantum aspects of gravity. On the phenomenological side, it is worth noting that the spherical mass accretion rate in power-Yang-Mills theories can be larger than its linear counterpart \cite{Gomez:2023qyv}. Additionally, the luminosity of accretion disks, i.e., the optical appearance and image formation, can also be affected, leading to a distinctive BH that can be probed in future observations of BH images \cite{Chakhchi:2022fls}.
\\
In a preliminary paper, we have investigated the impact of the non-Abelian charge on the BH properties and established certain relationships between the charges. 
In this paper, we further investigate the QNMs and shadow size of this type of BH, motivated by current observations, as discussed earlier. Specifically, the imaging observations provided by the EHT allow us to establish a more stringent constraint on the  non-Abelian charge. 
Concretely, our findings indicate that a slightly larger electric charge is allowed compared to the standard case for a Yang-Mills charge $Q_{\rm YM}\sim\mathcal{O}(-0.1)$, allowing the BH to maintain an event horizon.
This paper is structured as follows. In Section 2, we review the main elements of the model and discuss the key properties of the resulting BH. Within this section, we analyze the behavior of massless scalar fields propagating in the spherically symmetric gravitational background. We employ the WKB method and investigate the eikonal limit to gain further insights into the dynamics. Additionally, we calculate the size of the BH shadow and set bounds on both charges from imaging observations. We adopt the metric signature $-,+,+,+$, and work in geometrical units where the speed of light in vacuum and Newton's constant are set to unity, $G=1=c$.

\section{Background: Charged Black Hole solutions} \label{sec:2}

In this section, we will outline the key ingredients of the theory that leads to a novel non-linear BH solution. Our investigation is performed within a 4-dimensional spacetime, incorporating three crucial elements:
  i) The Einstein-Hilbert term,
 ii) The Maxwell invariant, and
iii) the Power Yang-Mills invariant.
Thus, the action that represents our scenario is:
\begin{align}
\begin{split}
     S_0 = \int \sqrt{-g}\ \mathrm{d}^4x 
     \Bigg[\frac{1}{2\kappa} &R - F_{\mu \nu} F^{\mu \nu}  -
  (F_{\mu \nu}^{(a)} F^{\mu \nu}_{(a)})^{p} 
     \Bigg].
     \label{action}
\end{split}
\end{align}
We have considered the usual definitions, namely:
  i) $G$ is Newton's constant,
 ii) $\kappa \equiv 8\pi G$ is Einstein's constant, 
iii) $g$ is the determinant of the metric tensor $g_{\mu \nu}$,
 iv) $R$ is the Ricci scalar, and
  v) $p$ is a real parameter that introduces non-linearities.
In addition, we have two extra tensors: i) the electromagnetic field strength $F_{\mu \nu}$, and ii)  the gauge strength tensor $F_{\mu \nu}^{(a)}$, both defined in terms of the potentials $A_\nu$ and $A_\nu^{(a)}$, respectively, and their corresponding expressions are:
\begin{align}
F_{\mu \nu} & \equiv \partial_\mu A_\nu - \partial_\nu A_\mu\,,
\\ 
F_{\mu \nu}^{(a)} & \equiv \partial_\mu A_\nu^{(a)} - \partial_\nu A_\mu^{(a)} + \frac{1}{2\sigma} C_{(b)(c)}^{(a)} 
A_{\mu}^{(b)} A_{\nu}^{(c)}\,.
\end{align}
It should be mentioned that Greek indices run from 0 to $3$ and $a$ is the internal gauge index running from $1$ to $3$.
Even more, $C_{(b)(c)}^{(a)}$ represents the structure constants of 3 parameter Lie group $\mathcal{G}$, 
$A^{(a)}_{\mu}$ are the $SO(3)$ gauge group Yang-Mills potentials, $\sigma$ is an arbitrary coupling constant, and finally $A_{\mu}$ is the conventional Maxwell potential.
At this point, it is essential to point out the concrete form of  $\mathbf{A}^{(a)}$ and $\mathbf{A}$. Thus, the first object is then defined as
\begin{align}
    \mathbf{A}^{(a)} = \frac{q_{\text{YM}}}{r^2}(x_i d x_j - x_j dx_i),
\end{align}
where $2 \le j+1	\le i 	\le 3$ and $1 \le a \le 3$. In addition, the radial coordinate is connected to $x_i$ according to $ r^2 = \sum_{i=1}^{3} x_i^2$.
The second object, the Maxwell potential 1-form, is therefore given by
\begin{align}
        \mathbf{A} &= \frac{Q}{r}dt,
\end{align}
where $Q$ represents the electric charge and $q_{\text{YM}}$ denotes the YM charge.
Varying the action with respect to the metric field we obtain Einstein's field equations, i.e., 
\begin{align}
    G_{\mu \nu} 
    &= 
    \kappa T_{\mu \nu},
\end{align}
where the energy-momentum tensor has two expected contributions: 
 i) the matter content, $T_{\mu \nu}^{\text{M}}$, and
ii) the Yang-Mills contribution, $T_{\mu \nu}^{\text{YM}}$, i.e., 
\begin{align}
    T_{\mu \nu} \equiv T_{\mu \nu}^{\text{M}} + T_{\mu \nu}^{\text{YM}}.
\end{align}
The last two contributions are defined in terms of $F_{\mu \nu}$ and $F_{\mu \nu}^{(a)}$ as follow
\begin{align}
    T_{\mu \nu}^{\text{M}} &= 2 F_{\mu}^{\lambda} F_{\nu \lambda} - \frac{1}{2} F_{\lambda \sigma} F^{\lambda \sigma} g_{\mu \nu}, 
    \\
    T_{\mu \nu}^{\text{YM}} &= -\frac{1}{2} g_{\alpha \mu}
    \Bigg[
    \delta^{\alpha}_{\nu} \mathcal{F}_{\text{YM}}^p - 4p \mathbf{Tr} 
    \Bigl(
    F_{\nu \lambda}^{(a)}F^{(a) \alpha \lambda}
    \Bigl)
    \mathcal{F}_{\text{YM}}^{p-1}
    \Bigg].
\end{align}
Varying the action with respect to the gauge potentials $\bf{A}$ and $\bf{A}^{(a)}$, we obtain the Maxwell and Yang Mills equations respectively
\begin{align}
  \mathrm{d}\Bigl( {} ^{\star} \mathbf{F}  \Bigl) &= 0,
\\
    \mathbf{d}\Bigl( {} ^{\star} \mathbf{F}^{(a)} \mathcal{F}_{\text{YM}}^{p-1}  \Bigl)   +  \frac{1}{\sigma} C^{(a)}_{(b)(c)} \mathcal{F}_{\text{YM}}^{p-1} \mathbf{A}^{(b)} \wedge^{\star} \mathbf{F}^{(c)} &= 0,
\end{align}
where $\star$ means duality. It is important to point out that the trace of the Yang-Mills gauge strength tensor takes the form:
\begin{align}
    \mathcal{F}_{\text{YM}} =  \frac{q^2_{\text{YM}}}{r^4},
\end{align}
which is positive, allowing us thus to consider all rational numbers for the $p$-values. It is evident that for $p = 1$, the formalism reduces to the standard Einstein Yang-Mills theory. 
It is even more obvious that when we consider $p=0$, the Yang-Mills term mimics the cosmological constant contribution, as can be easily checked. Thus, we then recover the Reissner-Nordstr\"{o}m-(anti) de Sitter metric.
In what follows, we consider a spherically symmetric space-time (in  Schwarzschild coordinates), and write the line element 
\begin{equation}
ds^{2} = - f(r) dt^{2} + f(r)^{-1} dr^{2} + r^{2}  (d\theta^2 + \sin^{2}\theta d\phi^{2}),
\label{lineelement}
\end{equation}
where $r$ is the radial coordinate. From the effective Einstein's field equations and the Maxwell and Yang-Mills equations we obtain
\begin{equation}
f(r)=1 - \frac{2M}{r} + \frac{Q^{2}}{r^{2}} + \frac{Q_{\rm YM}}{r^{4p-2}}.
\label{metricfunction}
\end{equation}
From the previous equation, we immediately notice that the Yang-Mills charge $q_{\rm YM}$ is related to its normalized version as follows \cite{Mazharimousavi:2009mb} 
\begin{align}
  Q_{\rm YM} &\equiv \frac{2^{p - 1}} {4 p - 3}   q_{\rm YM}^{2p} ,
\end{align}
for $p\neq3/4$.  The specific case of $p=3/4$ poses certain challenges due to the emergence of a radial logarithmic dependency in the solution, making it impossible to obtain an analytical solution. At this point, it should be noted that most studies in the pure Power Yang-Mills theory have focused on the range \(3/4 \leq p < 3/2\) because it aligns with all energy conditions of GR (see for instance \cite{Mazharimousavi:2009mb}). This stands in contrast to the case  $p=1/2$, which fails to meet some of the energy conditions. However, there exist classical and quantum field theories that can violate all energy conditions while remaining consistent with experimental data \cite{Flanagan:1996gw,Visser:1999de}. For example, violation of the energy conditions suggests the possible existence of traversal wormholes in theories with a non-minimally coupled scalar field with a positive curvature coupling \cite{Barcelo:2000zf}. Another example of fields violating the energy conditions are the inflaton and quintessence fields needed to explain early and late dynamics of the Universe. Moreover, in theories beyond GR, where additional degrees of freedom introduce new gravitational dynamics, the corresponding energy conditions assume a different character since the causal structure and geodesic structure may be altered \cite{Capozziello:2013vna}.  
Given these considerations, we adopt a more phenomenological perspective in this work, with no bias toward the case 
$p=1/2$. We defer a more technical study of these issues to future work, particularly in a more generalized scenario where both the Maxwell and Yang-Mills fields are non-trivially coupled\footnote{When Yang-Mills fields are coupled to other fields, other than the gravitational one, explicit interactions arise, leading to a more involved (effective) energy-momentum tensor of the Yang-Mills theory (see, e.g., \cite{Gomez:2021jbo}).}. Although we do not provide a formal proof at this stage, we maintain an agnostic stance on whether all energy conditions must universally hold in every physical situation. In this vein, the case 
$p=1/2$ warrants significant attention due to its simplicity and relevance in astrophysical contexts \cite{Gomez:2023qyv}. Unlike other explored cases, it modifies the structure of the Reissner-Nordstr\"{o}m spacetime in a non-trivial manner by introducing of a non-commutative charge into the solution \cite{Gomez:2023qyv}.
Even more, this case provides illuminating analytical solutions for the inner $r_{-}$ and external (event horizon) $r_{+}$ radii:
\begin{equation}
r\pm = \frac{M\pm \sqrt{M^{2}-Q^{2}(1+Q_{\rm YM})}}{1+Q_{\rm YM}}.
\label{horizons}
\end{equation}
We call this solution henceforth modified Reissner-Nordstr\"{o}m (MRN) solution with $Q_{\rm YM}\neq-1$. 
In addition, notice that for this concrete value of the power ($p=1/2$), $Q_{\rm YM}$ is a dimensionless parameter.
The precise form of the event horizon radius holds significant importance as it establishes a well-defined relationship between the two charges, thereby preventing the occurrence of a naked singularity,\footnote{The formation of a naked singularity in any gravitational theory is, however, not guaranteed by the vanishing of the horizon. Hence, a formal astrophysical collapse must be then carried out. This is of course beyond the scope of this paper.} among other astrophysical implications \cite{Gomez:2023qyv}.
It yields 
\begin{equation}
    Q_{\rm YM}>-1\; \text{and}\; 0<\frac{Q}{M}<\sqrt{\frac{1}{1+Q_{\rm YM}}}.\label{conditionsWCC}
\end{equation}
The conventional restriction for the Reissner-Nordstr\"{o}m  BH is covered in the previous expression and is consistently retrieved in the limit of $Q_{\rm YM} \rightarrow 0$, giving $Q/M < 1$ as it should be.
For the allowed range of values of both charges, the corresponding horizons $r_{+}$ are completely determined. 
%

\section{Wave equations and perturbations}

\subsection{Scalar perturbations}

\begin{figure*}[ht!]
\centering
\includegraphics[scale=0.95]{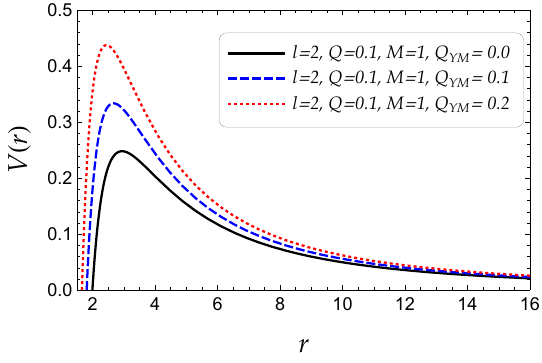} \
\includegraphics[scale=0.95]{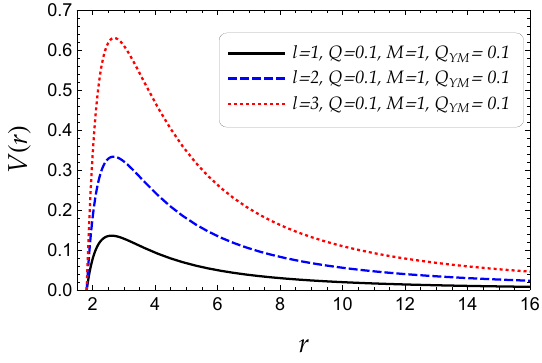} \
\\
\includegraphics[scale=0.95]{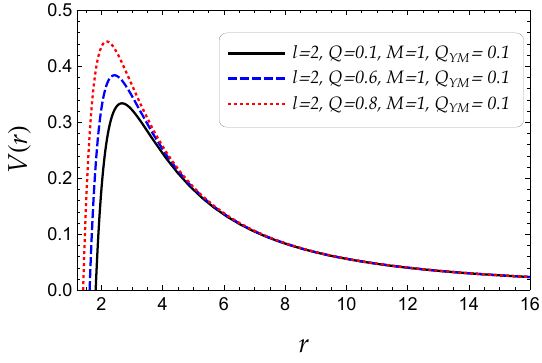} \
\includegraphics[scale=0.95]{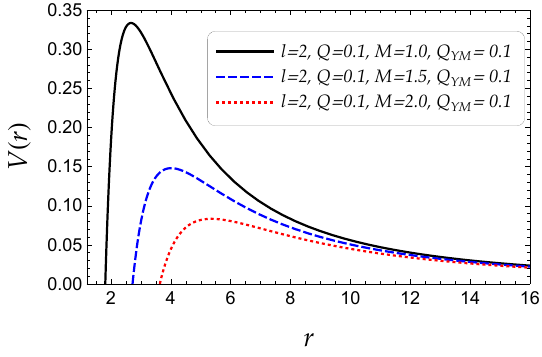} \
\caption{
Effective potential barrier for scalar perturbations against the  radial coordinate for the parameters shown in the panels. 
{\bf{Top Left panel:}} Effective potential for fixed values of $\{ \ell, Q, M \}$ and $ Q_{\text{YM}} = \{ 0.0, 0.1, 0.2 \}$.
{\bf{Top Right panel:}} Effective potential for fixed values of $\{ Q, M, Q_{\text{YM}} \}$ and $\ell = \{ 1, 2, 3 \}$.
{\bf{Bottom Left panel:}} Effective potential for fixed values of $\{ \ell, M, Q_{\text{YM}} \}$ and $Q = \{ 0.1, 0.6, 0.8 \}$.
{\bf{Bottom Right panel:}} Effective potential for fixed values of $\{ \ell, Q, Q_{\text{YM}} \}$ and $M = \{ 1.0, 1.5, 2.0 \}$.
}
\label{fig:1} 	
\end{figure*}

We begin by examining the propagation of a test scalar field, denoted as $\Phi$, in a fixed gravitational background within a four-dimensional spacetime. Additionally, we assume that the field is real. By considering the corresponding action $S[g_{\mu \nu}, \Phi]$, we can derive the following expression.
\begin{align}
S[g_{\mu \nu} ,\Phi] \equiv \frac{1}{2} \int \mathrm{d}^4 x \sqrt{-g}
\Bigl[
\partial^{\mu} \Phi \partial_{\mu} \Phi 
\Bigl]\,.
\end{align}
From here, we find the standard Klein-Gordon equation \cite{Crispino:2013pya,Kanti:2014dxa,Pappas:2016ovo,Panotopoulos:2019gtn,Avalos:2023ywb,Gonzalez:2022ote,Rincon:2020cos}
\begin{equation}
\frac{1}{\sqrt{-g}}\partial_{\mu}\left(\sqrt{-g}g^{\mu\nu}\partial_{\nu}\Phi\right) = 0.
\end{equation}
To decouple and eventually solve the Klein-Gordon equation, we take advantage of the symmetries of the metric and propose as an ansatz the following separation of variables in spherical coordinates as
\begin{equation}
\Phi(t, r, \theta, \phi) = e^{-i\omega t}\frac{\psi(r)}{r}Y_{\ell m}(\theta, \phi).
\end{equation}
Here, $Y_{\ell m}(\theta, \phi)$ represents the spherical harmonics, which solely depend on the angular coordinates. The quasinormal frequency, denoted as $\omega$, will be determined by selecting appropriate boundary conditions. Thus, the differential equation to be solved is:
\begin{align}
\begin{split}
& \frac{\omega^{2}r^{2}}{f(r)} + \frac{r}{\psi(r)}\frac{d}{dr}\left[r^{2}f(r)\frac{d}{dr}\left(\frac{\psi(r)}{r}\right)\right] +
\frac{1}{Y(\Omega)}\left[\frac{1}{\sin\theta}\frac{\partial}{\partial\theta}\left(\sin\theta\frac{\partial Y(\Omega)}{\partial\theta}\right)\right] +
\frac{1}{\sin^{2}\theta}\frac{1}{Y(\Omega)}\frac{\partial^{2}Y(\Omega)}{\partial\phi^{2}} = 0.
\label{KG}
\end{split}
\end{align}
The associated angular part can be recast as
\begin{align}
    \begin{split}
&\frac{1}{\sin\theta}\frac{\partial}{\partial\theta}\left(\sin\theta\frac{\partial Y(\Omega)}{\partial\theta}\right) + \frac{1}{\sin^{2}\theta}\frac{\partial^{2}Y(\Omega)}{\partial\phi^{2}} = 
-\ell(\ell + 1)Y(\Omega),
\end{split}
\end{align}
where $\ell(\ell + 1)$ is the corresponding eigenvalue, and $\ell$ is the angular degree. Combining the last two equations, we obtain a second-order differential equation for the radial coordinate. Now, considering the definition of  the "tortoise coordinate" $r_{*}$
\begin{align}
    r_{*}  \equiv  \int \frac{\mathrm{d}r}{f(r)}\,,
\end{align}
we can re-write the resulting differential equation in its Schr{\"o}dinger-like form, namely
\begin{equation} \label{SLE}
\frac{\mathrm{d}^{2}\psi(r_*)}{\mathrm{d}r_{*}^{2}} + \left[\omega^{2} - V(r_*)\right]\psi(r_*) = 0.
\end{equation}
Here $V(r)$ is the effective potential barrier defined as
\begin{equation}
V(r) = f(r)
\Bigg[ 
\frac{\ell(\ell + 1)}{r^{2}} + \frac{f'(r)}{r}
\Bigg],\label{poten}
\end{equation}
where the prime denotes derivative with respect to the radial variable.
Last but not least, the wave equation must be supplemented by appropriated
boundary conditions. In this case, such conditions are:
\begin{align}
   \Phi \rightarrow \: &\exp(+i \omega r_*), \; \; \; \; \; \;  r_* \rightarrow - \infty ,
   \\
   \Phi \rightarrow \: &\exp(-i \omega r_*), \; \; \; \; \; \; r_* \rightarrow + \infty .
\end{align}
Given the time dependence characterized by $\Phi \sim \exp(-i \omega t)$, a frequency with a negative imaginary part indicates a decaying (stable) mode. Conversely, a frequency with a positive imaginary part indicates an increasing (unstable) mode.
We show, in Fig.~\eqref{fig:1}, the behavior of the effective potential barrier $V(r)$ against the radial coordinate $r$ for different values of the set of parameters $\{ \ell, Q, M, Q_{\text{YM}} \}$. In all our analysis, we illustrate the Reissner-Nordstr\"{o}m case with $Q_{\rm YM}=0$ for comparison purposes.
Thus, from Fig.~\eqref{fig:1}, we can identify the following:
\begin{itemize}
    \item Top-Left panel shows the $V(r)$ for fixed $\{ \ell, Q, M \}$ and different values of the Yang-Mills charge $Q_{\text{YM}}$. 
    We observe that when $Q_{\text{YM}}$ increases, the maximum of the potential increases, at the time it shifts to the left. All solutions converge at small radii because their associated horizons are equals in contract to the other cases depicted.
    \item Top-Right panel shows the $V(r)$ for fixed $\{ Q, M, Q_{\text{YM}} \}$ and different values of the angular degree $\ell$. 
    We observe that when $\ell$ increases, the maximum of the potential increases, shifting equally to the right.
    \item Bottom-Left panel shows the $V(r)$ for fixed $\{ \ell, M, Q_{\text{YM}} \}$ and different values of the charge $Q$.
    We observe that when $Q$ increases, the maximum of the potential increases, at the time it shifts to the left. In addition, the potential tends to be overlapped for moderated and large values of $r$. 
    \item Bottom-Right panel shows the $V(r)$ for fixed $\{ \ell, Q, Q_{\text{YM}} \}$ and different values of the BH mass $M$. 
    We observe that when $M$ increases, the maximum of the potential decreases, and the potential shifts significantly to the right compared to the other panels.
\end{itemize}

\subsection{Dirac perturbations}
In what follows, we will summarize the basic expression to compute QNMs for neutral Dirac particles. In doing so, it is convenient to remember the vierbein formalism and, subsequently, the corresponding effective potential, following \cite{Destounis:2018qnb}.
Let us start by introducing the tetrad (or vierbein), $e_a^\mu$, defined according to
\begin{equation}
e_\mu^a e_\nu^b \eta_{ab} = g_{\mu \nu} ,
\end{equation}
where $\eta_{ab}$ is the flat Minkowski metric tensor. The vierbein carries two kinds of indices: 
 i) a flat index $a$, and 
ii) a space-time index $\mu$, 
and it could be considered as the "square root" of the metric tensor $g_{\mu \nu}$. 
Next, we define curved Dirac matrices as follows
\begin{equation}
G^\mu \equiv e_a^\mu \gamma^a ,
\end{equation}
which have the following properties
\begin{eqnarray}
\{ \gamma^a, \gamma^b \} & = & -2 \eta^{ab}, \\
\{ G^\mu, G^\nu \} & = & -2 g^{ab},
\end{eqnarray}
Here, $\gamma^a$ are the conventional Dirac matrices from  relativistic quantum mechanics (in flat space-time). 
To conclude this summary, let us introduce the spin connection  $\omega_{ab \mu}, \Gamma_\mu$
\begin{eqnarray}
\Gamma_\mu & = & -\frac{1}{8} \omega_{ab\mu} [\gamma^a, \gamma^b] \\
\omega_{ab \mu} & = & \eta_{ac} [e_\nu^c e_b^\lambda \Gamma^\nu_{\mu \lambda} - e_b^\lambda \partial_\mu e_\lambda^c] ,
\end{eqnarray}
where $\Gamma^\nu_{\mu \lambda}$ are the Christoffel symbols.
Finally, the Dirac equation for a spin one-half fermion $\Psi$ in curved space-time is 
\begin{equation}
(i G^\mu D_\mu - m_f) \Psi = 0.
\end{equation}
Note that; 
i) $m_f$ is the mass of the fermion, and
ii)  $D_\mu \equiv \partial_\mu + \Gamma_\mu$ is the covariant derivative.
For Dirac fermions, we separate variables as before, where now we use the spinor spherical harmonics \cite{Finster:1998ak}, and two radial parts
\begin{align}
r^{-1} \: f(r)^{-1/4} \: F(r) ,
\\
r^{-1} \: f(r)^{-1/4} \: i G(r) ,
\end{align}
for the upper and lower components of the Dirac spinor $\Psi$, respectively. 
In the massless limit, $m_f=0$, we obtain the following equations \cite{Destounis:2018qnb}
\begin{eqnarray}
\frac{dF}{dx} - W F + \omega G & = & 0 , \\
\frac{dG}{dx} + W G - \omega F & = & 0 ,
\end{eqnarray}
where $x$ is the tortoise coordinate as before, while the function $W$ is given by \cite{Destounis:2018qnb}
\begin{equation}
W = \frac{\xi \sqrt{f}}{r} ,
\end{equation}
where $\xi = \pm (j+1/2)=\pm 1, \pm 2,...$, with $j$ being the total angular momentum, $j=l \pm 1/2$ \cite{Destounis:2018qnb}. Finally, the two coupled equations for $F,G$ can be combined to obtain a wave equation for each one of the following form
\begin{eqnarray}
\frac{d^2F}{dx^2} + [\omega^2 - V_+] F & = & 0 ,\\
\frac{d^2G}{dx^2} + [\omega^2 - V_-] G & = & 0 ,
\end{eqnarray}
where the potentials are given by 
\begin{equation}
V_{\pm} = W^2 \pm \frac{dW}{dx} .
\end{equation}
In the framework of supersymmetry, the potentials \(V_-\) and \(V_+\) are considered superpartners, derived from a superpotential. Consequently, they yield identical spectra (see, e.g., \cite{Cooper:1994eh}). Thus, we will proceed using the plus sign and the wave equation for \(F(r)\).
To simplify the work, the effective potential in standard radial coordinates acquire the simple form:
\begin{align}
V_{+} &=   f(r) \left(\frac{\xi  f'(r)}{2 r \sqrt{f(r)}}-\frac{\xi  \sqrt{f(r)}}{r^2}+\frac{\xi ^2}{r^2}\right),
\end{align}
or, replacing the metric potential explicitly, by 
\begin{align}
    V_{+} &= \left(-\frac{2 M}{r}+\frac{Q^2}{r^2} + Q_{\text{YM}}+1\right) 
    \left( 
    - \frac{\xi}{r^2}  \sqrt{-\frac{2 M}{r}+\frac{Q^2}{r^2}+Q_{\text{YM}}+1} 
    + \frac{\xi  \left(\frac{2 M}{r^2}-\frac{2 Q^2}{r^3}\right)}{2 r \sqrt{-\frac{2 M}{r}+\frac{Q^2}{r^2} + Q_{\text{YM}} +1}}
    + \frac{\xi ^2}{r^2}
    \right).
\end{align}
\begin{figure*}[ht!]
\centering
\includegraphics[scale=0.95]{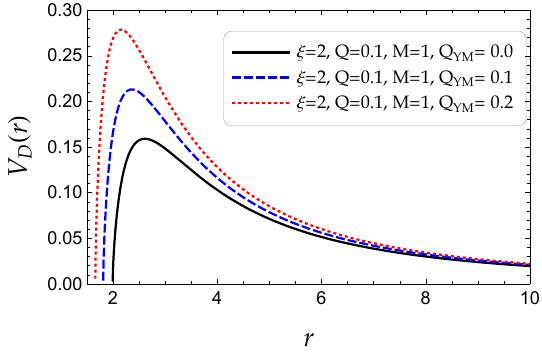} \
\includegraphics[scale=0.95]{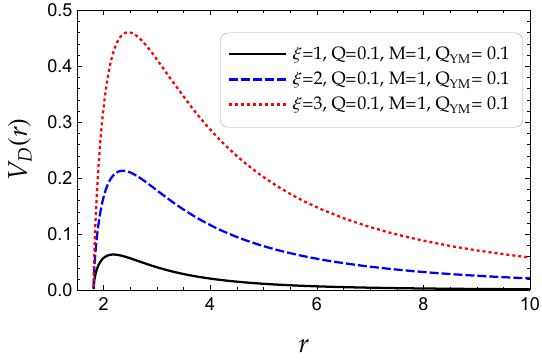} \
\\
\includegraphics[scale=0.95]{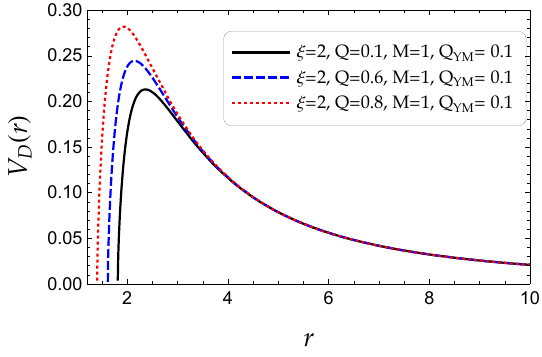} \
\includegraphics[scale=0.95]{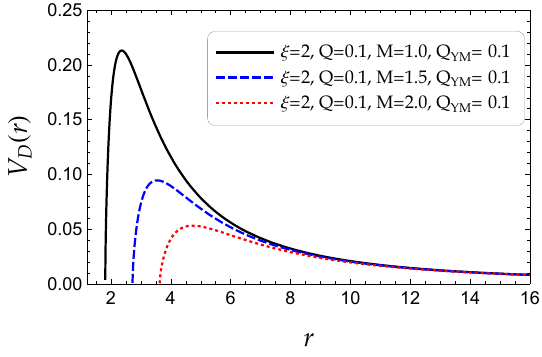} \
\caption{
Effective potential barrier for Dirac perturbations against the radial coordinate for the parameters shown in the panels. 
{\bf{Top Left panel:}} Effective potential for fixed values of $\{ \xi, Q, M \}$ and $ Q_{\text{YM}} = \{ 0.0, 0.1, 0.2 \}$.
{\bf{Top Right panel:}} Effective potential for fixed values of $\{ Q, M, Q_{\text{YM}} \}$ and $\xi = \{ 1, 2, 3 \}$.
{\bf{Bottom Left panel:}} Effective potential for fixed values of $\{ \xi, M, Q_{\text{YM}} \}$ and $Q = \{ 0.1, 0.6, 0.8 \}$.
{\bf{Bottom Right panel:}} Effective potential for fixed values of $\{ \xi, Q, Q_{\text{YM}} \}$ and $M = \{ 1.0, 1.5, 2.0 \}$.
}
\label{fig:1D} 	
\end{figure*}

As for the scalar case, the wave equation is completed with some adequate boundary conditions, which are
\begin{align}
   \Phi \rightarrow \: &\exp(+i \omega r_*), \; \; \; \; \; \;  r_* \rightarrow - \infty ,
   \\
   \Phi \rightarrow \: &\exp(-i \omega r_*), \; \; \; \; \; \; r_* \rightarrow + \infty .
\end{align}
Similarly, as $\Phi \sim \exp(-i \omega t)$, a frequency with a negative imaginary part reveals a decaying (stable) mode, and a frequency with a positive imaginary part indicates an increasing (unstable) mode.
To exemplify how the effective potential $V_D(r)$ looks like in the Dirac case, we depict, in Fig.~\eqref{fig:1D}, its behavior against the radial coordinate for different values of the set of parameters $\{ \xi, Q, M, Q_{\text{YM}} \}$. We include the Reissner-Nordstr\"{o}m case with $Q_{\rm YM}=0$ for comparison purposes.
We have considered in Fig.~\eqref{fig:1D} the following cases:
\begin{itemize}
    \item Top-Left panel shows the $V_D(r)$ for fixed $\{ \xi, Q, M \}$ and different values of the Yang-Mills charge $Q_{\text{YM}}$. 
    We noticed that when $Q_{\text{YM}}$ increases, the maximum of the potential increases, at the time it shifts to the left. All solutions converge at small radii because their associated horizons are equals in contract to the other cases depicted.
    \item Top-Right panel represents the $V_D(r)$ for fixed $\{ Q, M, Q_{\text{YM}} \}$ and different values of the angular parameter $\xi$. 
    The figure reveals that when $\xi$ increases, the maximum of the potential increases, shifting now to the right.
    \item Bottom-Left panel shows the $V_D(r)$ for fixed $\{ \xi, M, Q_{\text{YM}} \}$ and different values of the charge $Q$.
    The figure confirms that when $Q$ increases, the maximum of the potential increases, at the time it shifts to the left. Moreover, the effective potential tends to overlap quickly from small radii.
    \item Bottom-Right panel depicts the $V_D(r)$ for fixed $\{ \xi, Q, Q_{\text{YM}} \}$ and various values of the BH mass $M$. 
    The figure shows that when $M$ increases, the maximum of the potential decreases, and the potential shifts significantly to the right compared to the other panels.
\end{itemize}

\section{Quasinormal spectrum}

\subsection{WKB approximation}

Exact analytical expressions for the quasinormal spectra of BHs can only be obtained in a limited number of cases. For example:
i) When the effective potential barrier takes the form of the P{\"o}schl-Teller potential, as studied in references such as \cite{Poschl:1933zz,Ferrari:1984zz,Cardoso:2001hn,Cardoso:2003sw, Molina:2003ff,Panotopoulos:2018hua}.
ii) When the corresponding differential equation for the radial part of the wave function can be transformed into the Gauss' hypergeometric function, as explored in references  \cite{Birmingham:2001hc,Fernando:2003ai,Fernando:2008hb,Gonzalez:2010vv,Destounis:2018utr,Ovgun:2018gwt,Rincon:2018ktz}.
Considering the complexity and non-trivial nature of the involved differential equation, it becomes necessary to rely on numerical or, at the very least, semi-analytical methods to compute the corresponding quasinormal frequencies. Consequently, numerous techniques have been developed for this purpose, some of which are commonly utilized. Specifically:
i) The Frobenius method and its generalization, as referenced in \cite{Destounis:2020pjk, Fontana:2022whx,Hatsuda:2021gtn}.
ii) The method of continued fraction, along with its enhancements, is mentioned in \cite{Leaver:1985ax,Nollert:1993zz, Daghigh:2022uws}. 
iii) The asymptotic iteration method \cite{Cho:2011sf,2003JPhA...3611807C,Ciftci:2005xn}
among others.
Additional details can be found in \cite{Konoplya:2011qq} for more comprehensive information.
In the present paper, we will implement the well-known WKB semi-classical method to obtain the quasinormal frequencies (see \cite{Schutz:1985km,Iyer:1986np,Iyer:1986nq,Kokkotas:1988fm,Seidel:1989bp} for technical details).
The WKB method is a commonly used semi-analytic approach for computing the QNMs of BHs. The initial first-order computation was derived by Schutz and Will \cite{Schutz:1985km}, followed by subsequent improvements made by Iyer and Will \cite{Iyer:1986np}, who developed a semi-analytic formula incorporating second and third-order corrections. This method has demonstrated remarkable efficiency in determining the lowest overtones among the complex frequencies of an oscillating Schwarzschild BH. The accuracy of the approximation improves with increasing values of the angular harmonic index $\ell$, but deteriorates as the overtone index increases.
Building upon these advancements, R.A. Konoplya extended the generalization up to the 6th order \cite{Konoplya:2003ii}, while J. Matyjasek and M. Opala found the formulae from the 7th to the 13th order \cite{Matyjasek:2017psv}.
It should be emphasized that the higher-order equations of the WKB method have not been mathematically proven to consistently converge to the theoretical resolution. Some years ago, R. Konoplya introduced a preliminary criterion for identifying errors, based on the simple subtraction of frequencies at successive orders in the WKB method. Although this idea partially works, it does not provide a strict criterion. Consequently, depending on the complexity of the BH spacetime, the higher-order results obtained by the WKB method may show certain deviations and errors. In cases with simpler backgrounds, such as those considered in this manuscript, the WKB method is capable of producing accurate results. Therefore, for finding quasinormal (QN) frequencies with high precision, this method is often sufficient, and there may be no need to compare it with alternative approaches.

The method relies on the resemblance of \eqref{SLE} to the one-dimensional Schrodinger equation corresponding to a potential barrier.
The WKB formula employs the matching of asymptotic solutions, which consist of a combination of ingoing and outgoing waves, along with a Taylor expansion centered around the peak of the potential barrier at $x=x_0$. This expansion encompasses the region between the two turning points, which correspond to the roots of the effective potential $U(x,\omega) \equiv V(x) - \omega^2$.
In what follows, we will implement the WKB method to compute the QN spectra of 6th order, by means of the following expression
\begin{equation}
\omega_n^2 = V_0+(-2V_0'')^{1/2} \Lambda(n) - i \nu (-2V_0'')^{1/2} [1+\Omega(n)]\,,
\end{equation}
where 
i) $V_0''$ represents the second derivative of the potential at the maximum, 
ii) $\nu = n+1/2$, 
iii) $V_0$ symbolizes the maximum of the effective barrier, and 
iv) $n=0,1,2...$ is the overtone number.
In addition, $\Lambda(n), \Omega(n)$ are long and intricate relations of $\nu$ (and derivatives of the potential evaluated at the maximum), the reason why we avoid to show the concrete form of them. Instead, they can be found, for instance, in \cite{Kokkotas:1988fm}. 
Thus, to perform our computations, we have used here a Wolfram Mathematica \cite{wolfram} notebook utilizing the WKB method at any order from one to six \cite{Konoplya:2019hlu}.
In addition, for a given angular degree, $\ell$, we will consider values $n < \ell$ only. For higher order WKB corrections (and recipes for simple, quick, efficient and accurate computations) see \cite{Konoplya:2019hlu,Hatsuda:2019eoj}. 
Finally, notice that as was pointed out (for instance by R. Konoplya \cite{Konoplya:2019hlu}), the WKB series converges asymptotically only, there is no mathematically strict criterion for the evaluation of an error. However, the sixth/seventh order usually produces the best result. 
Similarly, it has been verified that for low values of $\ell$ (while keeping $n < \ell$), the quasinormal frequencies obtained using the WKB method still yield quite accurate results \cite{Konoplya:2004ip}.

\begin{figure*}[ht!]
\centering
\includegraphics[scale=0.95]{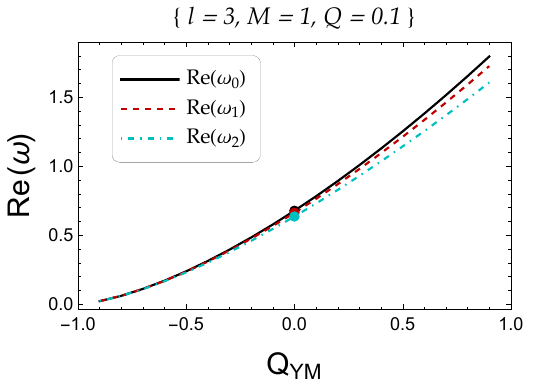} \
\includegraphics[scale=0.95]{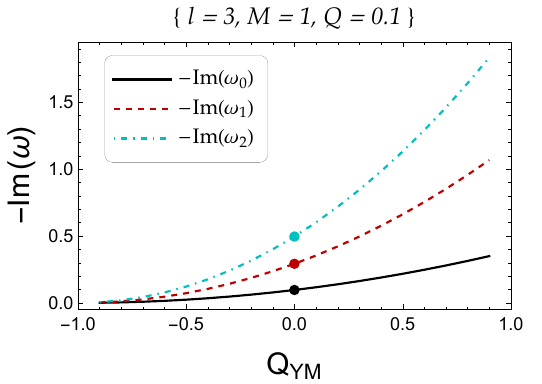} \
\\
\includegraphics[scale=0.95]{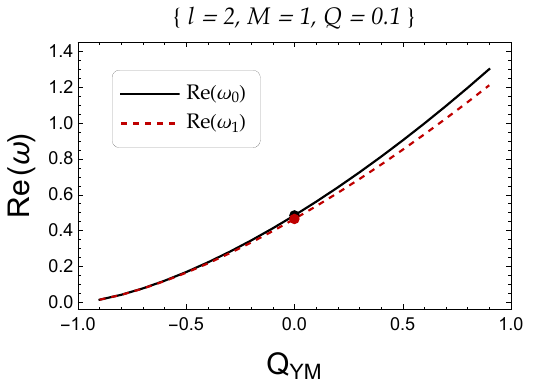} \
\includegraphics[scale=0.95]{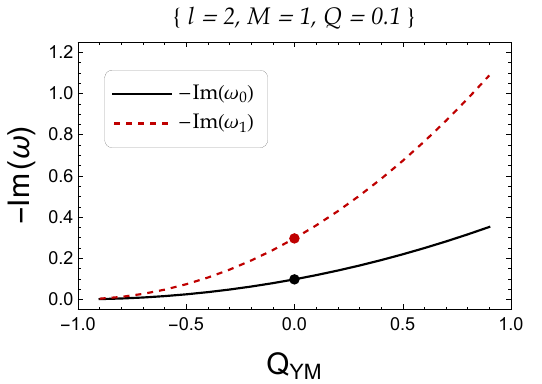} \
\\
\includegraphics[scale=0.95]{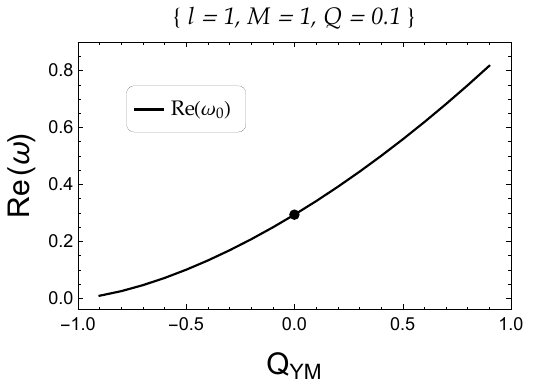} \
\includegraphics[scale=0.95]{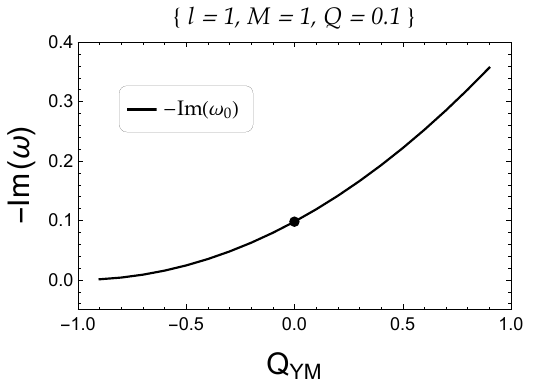} \
\caption{
QNMs for all cases investigated with $M=1$, $\ell= \{3, 2, 1 \}$ and $n=\{ 0, 1 \}$. 
{\bf{Left column}}: Real part of $\omega$ against the Yang-Mills charge $Q_{\text{YM}}$.
{\bf{Right column}}: Imaginary part of $\omega$ against the Yang-Mills charge $Q_{\text{YM}}$.
The color code is:
  i) solid black line for $n=0$ and
 ii) dashed red line for $n=1$. 
iii) dot-dashed cyan line $n=2$.
We have assumed $Q=0.1$.
The Reissner-Nordstr\"{o}m case ($Q_{\text{YM}}=0$) is denoted with black, red, and cyan dots for comparison.
}
\label{fig:1A} 	
\end{figure*}

\begin{figure*}[ht!]
\centering
\includegraphics[scale=0.635]{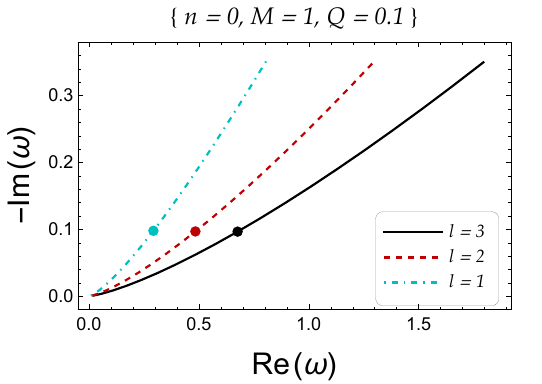} \
\includegraphics[scale=0.635]{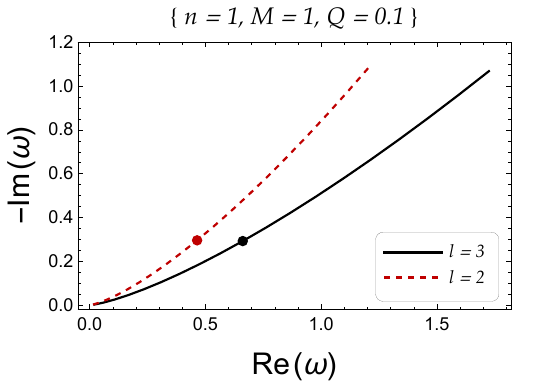} \
\includegraphics[scale=0.635]{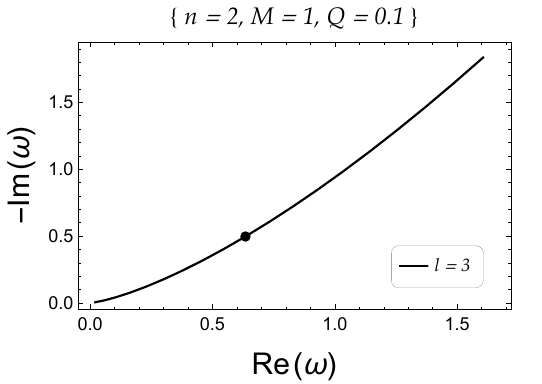} \
\caption{QNMs for all cases investigated with $M=1$, $\ell= \{3, 2, 1 \}$ and $Q=0.1$. 
The figures show the negative imaginary part of the frequency against the real part of the frequency for 
  i)  $n=0$, left panel
 ii)  $n=1$, middle panel, and 
iii)  $n=2$, right panel.
The Reissner-Nordstr\"{o}m case ($Q_{\text{YM}}=0$) is denoted with black, red, and cyan dots for comparison.}
\label{fig:2} 	
\end{figure*}

\begin{figure*}[ht!]
\centering
\includegraphics[scale=0.95]{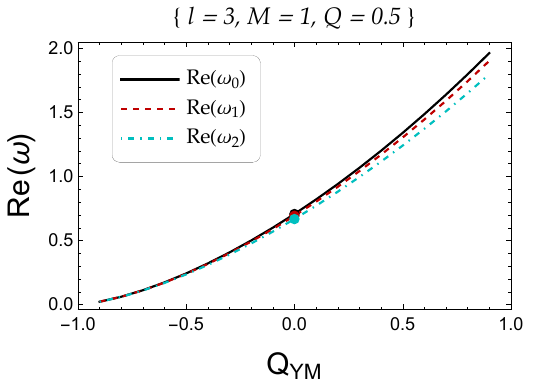} \
\includegraphics[scale=0.95]{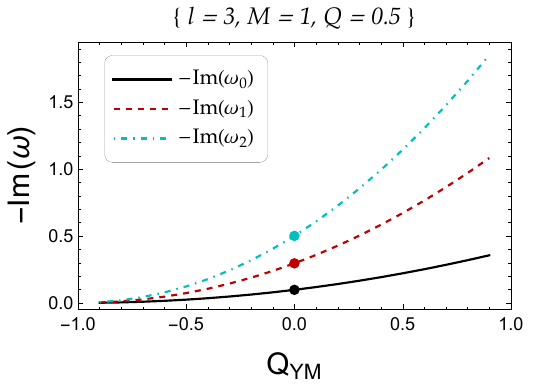} \
\\
\includegraphics[scale=0.95]{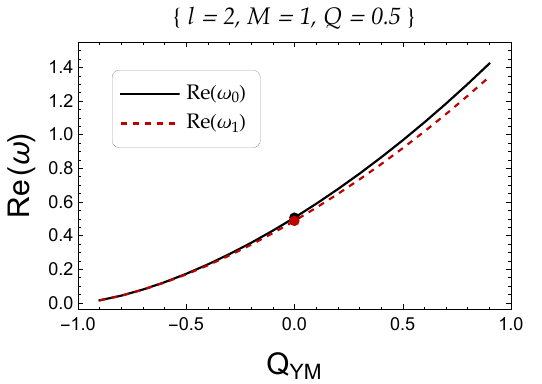} \
\includegraphics[scale=0.95]{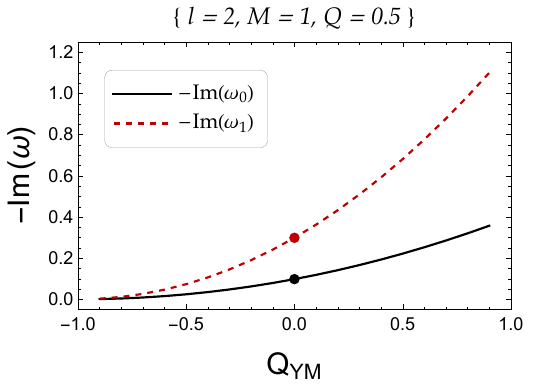} \
\\
\includegraphics[scale=0.95]{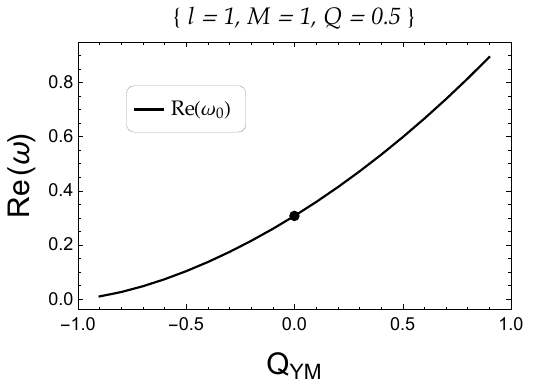} \
\includegraphics[scale=0.95]{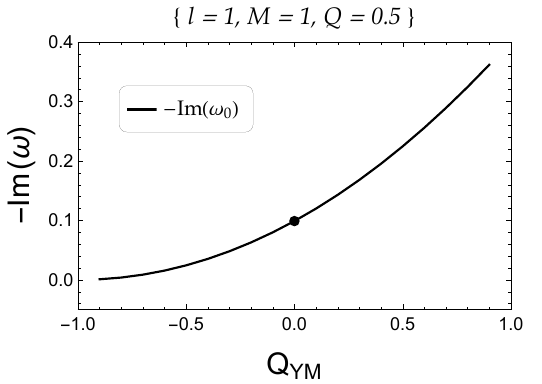} \
\caption{
QNMs for all cases investigated with $M=1$, $\ell= \{3, 2, 1 \}$ and $n=\{ 0, 1 \}$. 
{\bf{Left column}}: Real part of $\omega$ against the Yang-Mills charge $Q_{\text{YM}}$.
{\bf{Right column}}: Imaginary part of $\omega$ against the Yang-Mills charge $Q_{\text{YM}}$.
The color code is:
  i) solid black line for $n=0$ and
 ii) dashed red line for $n=1$. 
iii) dot-dashed cyan line $n=2$.
We have assumed $Q=0.5$.
The Reissner-Nordstr\"{o}m case ($Q_{\text{YM}}=0$) is denoted with black, red, and cyan dots for comparison.
}
\label{fig:5} 	
\end{figure*}

\begin{figure*}[ht!]
\centering
\includegraphics[scale=0.635]{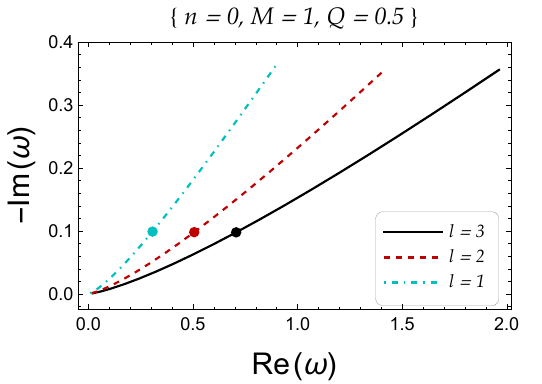} \
\includegraphics[scale=0.635]{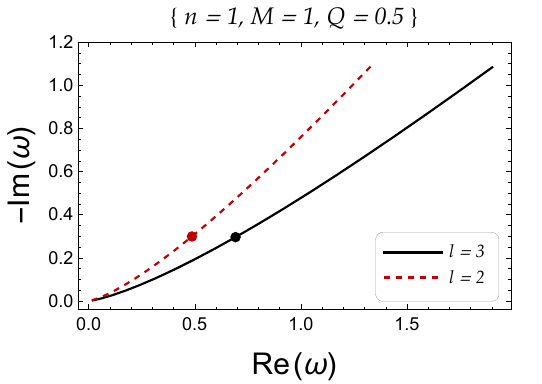} \
\includegraphics[scale=0.635]{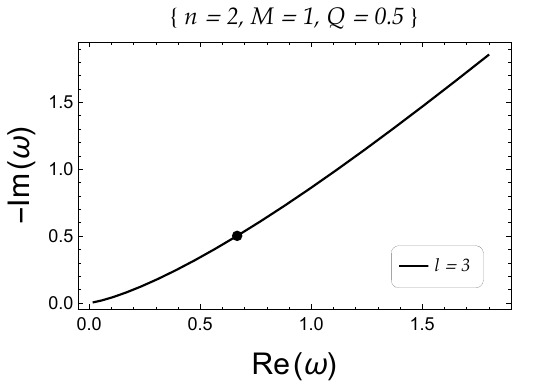} \
\caption{QNMs for all cases investigated with $M=1$, $\ell= \{3, 2, 1 \}$ and $Q=0.5$. 
The figures show the negative imaginary part of the frequency against the real part of the frequency for 
  i)  $n=0$, left panel
 ii)  $n=1$, middle panel, and 
iii)  $n=2$, right panel.
The Reissner-Nordstr\"{o}m case ($Q_{\text{YM}}=0$) is denoted with black, red, and cyan dots for comparison.
}
\label{fig:6} 	
\end{figure*}

For scalar perturbations, we summarize our results in figures \eqref{fig:1A},\eqref{fig:2},\eqref{fig:5} and \eqref{fig:6}, as well as tables \eqref{table:First set} and \eqref{table:Third set}, where the frequencies have been calculated numerically for different angular degrees $\ell=1,2,3$ using the WKB approximation of {\bf{6th order}}. To ensure the reliability of our results, we have included a comparison with the lower orders of the WKB approximation in the appendix. This comparison validates that the sixth order is the most appropriate for the range of parameters considered.  Based on the outcomes obtained through the WKB approximation, our results indicate the stability of all modes for the given numerical values. This feature will be supported through an alternative approach, namely the eikonal approximation. 

For Dirac perturbations, we summarize our results in table \eqref{table:Fifth set} varying $n$, $Q_{\text{YM}}$ and $\xi$. We computed the QN frequencies using the WKB method of {\bf{6th order}}. According to our numerical results, all modes are found to be stable (given the negative value of the quasinormal frequencies). It should be mentioned that the modes found are valid because they respect the condition $n<\xi$, a sector in which the WKB method works better.

\begin{table}[ph!]
\centering
\caption{Quasinormal frequencies (varying $\ell$, $n$ and $Q_{\text{YM}}$) with $M=1$ and $Q=0.1$ for the  model considered in this work. For comparison, we have highlighted the Reissner-Nordstr\"{o}m case in bold ($Q_{\text{YM}}=0$).
}
{
\begin{tabular}{c|c|ccc} 
\toprule
$Q_{\text{YM}}$ & $n$ &  $\ell=1$ & $\ell=2$ & $\ell=3$ 
\\ \colrule
     & 0 &  0.0086756 - 0.000963972 I & 0.0149487 - 0.000962866 I & 0.0211130 - 0.000962586 I \\
-0.9 & 1 &   				   		  & 0.0148804 - 0.002894360 I & 0.0210645 - 0.002890650 I  \\
     & 2 &  	 				   		&  			     		  & 0.0209681 - 0.004827370 I
\\ \botrule
     & 0 &  0.0247323 - 0.0038626 I & 0.0423986 - 0.0038539 I & 0.0598047 - 0.00385168 I \\
-0.8 & 1 &   				   		& 0.0420159 - 0.0116074 I & 0.0595314 - 0.01157810 I  \\
     & 2 &  	 				   	&  			     		  & 0.0589934 - 0.01937310 I
\\ \botrule
     & 0 &  0.0457903 - 0.00870564 I & 0.0781065 - 0.00867671 I  & 0.110030 - 0.00866926 I \\
-0.7 & 1 &  			   		     & 0.0770624 - 0.02618310 I  & 0.109281 - 0.02608530 I  \\
     & 2 &  	 				   	 &  			     		 & 0.107817 - 0.04373080 I
\\ \botrule
     & 0 &   0.0710409 - 0.0155026 I  & 0.120584 - 0.0154348 I  & 0.169651 - 0.0154173 I \\
-0.6 & 1 &   				   	   	  & 0.118462 - 0.0466639 I  & 0.168121 - 0.0464348 I  \\
     & 2 &  	 				   	  &  			     		& 0.165156 - 0.0779912 I
\\ \botrule
     & 0 &   0.100036 - 0.0242629 I  & 0.168983 - 0.0241318 I & 0.237442 - 0.0240978 I \\
-0.5 & 1 &   				  	     & 0.165313 - 0.0730911 I & 0.234783 - 0.0726488 I  \\
     & 2 &  	 				   	 &  			     	  & 0.229672 - 0.1222420 I
\\ \colrule
\hline
     & 0 &  0.132484 - 0.0349955 I  & 0.222740 - 0.034771 I   & 0.312583 - 0.0347126 I \\
-0.4 & 1 &  	  				  	& 0.217007 - 0.105504 I   & 0.308409 - 0.1047490 I  \\
     & 2 &  			   			&  			     	      & 0.300454 - 0.1765700 I
\\ \colrule
\hline
     & 0 &   0.168181 - 0.047709 I      & 0.281447 - 0.0473559 I  & 0.394475 - 0.0472636 I \\
-0.3 & 1 &  				  		    & 0.273103 - 0.1439420 I  & 0.388370 - 0.1427580 I \\
     & 2 &  				   			&  			     	      & 0.376830 - 0.2410550 I
\\ \colrule
\hline
     & 0 &   0.206975 - 0.0624116 I & 0.344792 - 0.0618898 I  & 0.482659 - 0.0617529 I \\
-0.2 & 1 &   				   		& 0.333256 - 0.1884430 I  & 0.474177 - 0.1866950 I  \\
     & 2 &    				   		&  			              & 0.458280 - 0.3157790 I
\\ \colrule
\hline
     & 0 &  0.248747 - 0.0791103 I     & 0.412529 - 0.0783763 I  & 0.576767 - 0.0781822 I \\
-0.1 & 1 &      			  		   & 0.397196 - 0.2390410 I  & 0.565439 - 0.2365830 I  \\
     & 2 &  				  		   &  			             & 0.544386 - 0.4008180 I
\\ \colrule
            & {\bf{0}} &   {\bf{0.293407 - 0.0978114 I}}  & {\bf{0.484455 - 0.0968185 I}} & {\bf{0.676499 - 0.0965534 I}} \\
 {\bf{0.0}} & {\bf{1}} &      			  		         & {\bf{0.464698 - 0.2957730 I}} & {\bf{0.661832 - 0.2924410 I}}  \\
            & {\bf{2}} &  				  		          &  			                  & {\bf{0.634804 - 0.4962460 I}}
\\ \colrule
\hline
    & 0 &  0.340877 - 0.118520 I   & 0.560403 - 0.117220 I  & 0.781601 - 0.116869 I \\
0.1 & 1 &      			  		   & 0.535576 - 0.358672 I  & 0.763082 - 0.354289 I  \\
    & 2 &  				  		   &  			            & 0.729246 - 0.602136 I
\\ \colrule
\hline
    & 0 &   0.391099 - 0.141240 I   & 0.640230 - 0.139584 I  & 0.891858 - 0.139129 I \\
0.2 & 1 &   				   		& 0.609672 - 0.427770 I  & 0.868955 - 0.422148 I  \\
    & 2 &    				   		&  			             & 0.827468 - 0.718554 I
\\ \colrule
\hline
    & 0 &   0.444021 - 0.165975 I   & 0.723814 - 0.163913 I  & 1.007080 - 0.163338 I \\
0.3 & 1 &  				  		    & 0.686851 - 0.503099 I  & 0.979248 - 0.496037 I  \\
    & 2 &  				   			&  			     		 & 0.929258 - 0.845569 I
\\ \colrule
\hline
    & 0 &  0.499599 - 0.192726 I    & 0.811049 - 0.190212 I  & 1.12711 - 0.189495 I \\
0.4 & 1 &  	  				  	  	 & 0.766996 - 0.584689 I  & 1.09378 - 0.575975 I  \\
    & 2 &  			   			  	 &  			     	   & 1.03443 - 0.983244 I
\\ \colrule
\hline
    & 0 &   0.557803 - 0.221494 I    & 0.901842 - 0.218482 I & 1.2518 - 0.217604 I \\
0.5 & 1 &   				  	    & 0.850004 - 0.67257 I & 1.2124 - 0.661981 I  \\
    & 2 &  	 				   		&  			     		 & 1.14284 - 1.13164 I
\\ \colrule
\hline
    & 0 &   0.618597 - 0.252281 I  & 0.996109 - 0.248728 I  & 1.38103 - 0.247667 I \\
0.6 & 1 &   				   		& 0.935783 - 0.766769 I  & 1.33496 - 0.754074 I  \\
    & 2 &  	 				   		&  			     		 & 1.25433 - 1.29082 I
\\ \botrule
    & 0 &   0.681956 - 0.285086 I   & 1.09378 - 0.280953 I  & 1.51467 - 0.279684 I \\
0.7 & 1 &   				   		& 1.02426 - 0.867312 I  & 1.46133 - 0.852272 I  \\
    & 2 &  	 				   		&  			     		& 1.36878 - 1.46084 I
\\ \botrule
    & 0 &   0.747864 - 0.319906 I   & 1.19478 - 0.315159 I  & 1.65262 - 0.313657 I \\
0.8 & 1 &   				   		& 1.11535 - 0.974224 I  & 1.5914 - 0.956592 I  \\
    & 2 &  	 				   		&  			     		  & 1.48609 - 1.64174 I
\\ \botrule
    & 0 &   0.816295 - 0.356742 I   & 1.29906 - 0.351349 I  & 1.79479 - 0.34959 I \\
0.9 & 1 &   				   		& 1.20900 - 1.087530 I  & 1.72506 - 1.06705 I  \\
    & 2 &  	 				   		&  			     		& 1.60616 - 1.83360 I
\\ \botrule
\end{tabular} 
\label{table:First set}
}
\end{table}


\begin{table}[ph!]
\centering
\caption{
Quasinormal frequencies (varying $\ell$, $n$ and $Q_{\text{YM}}$) with $M=1$ and $Q=0.5$ for the  model considered in this work. 
For comparison, we have highlighted the Reissner-Nordstr\"{o}m case in bold ($Q_{\text{YM}}=0$).
}
{
\begin{tabular}{c|c|ccc} 
\toprule
$Q_{\text{YM}}$ & $n$ &  $\ell=1$ & $\ell=2$ & $\ell=3$ 
\\ \colrule
     & 0 &  0.00871068 - 0.000965239 I & 0.0150091 - 0.000964138 I & 0.0211983 - 0.000963859 I \\
-0.9 & 1 &   				   		   & 0.0149411 - 0.002898150 I & 0.0211499 - 0.002894460 I  \\
     & 2 &  	 				   		  &  			     		  & 0.0210540 - 0.004833670 I
\\ \botrule
     & 0 &  0.0249345 - 0.0038726 I & 0.0427446 - 0.00386398 I & 0.0602924 - 0.00386178 I \\
-0.8 & 1 &   				   		& 0.0423655 - 0.01163720 I & 0.0600217 - 0.01160820 I  \\
     & 2 &  	 				   	   &  			     		  & 0.0594888 - 0.01942250 I
\\ \botrule
     & 0 &  0.0463581 - 0.00873895 I & 0.0790722 - 0.00871039 I  & 0.111389 - 0.00870304 I \\
-0.7 & 1 &  			   		     & 0.0780432 - 0.02628180 I  & 0.110651 - 0.02618540 I  \\
     & 2 &  	 				        &  			     		    & 0.109209 - 0.04389380 I
\\ \botrule
     & 0 &   0.0722284 - 0.0155804 I  & 0.122592 - 0.0155138 I  & 0.172474 - 0.0154966 I \\
-0.6 & 1 &   				   	   	  & 0.120512 - 0.0468933 I  & 0.170973 - 0.0466689 I  \\
     & 2 &  	 				   	  &  			     		   & 0.168067 - 0.0783687 I
\\ \botrule
     & 0 &   0.102149 - 0.0244122 I  & 0.172537 - 0.0242842 I & 0.242430 - 0.0242510 I \\
-0.5 & 1 &   				  	     & 0.168959 - 0.0735298 I & 0.239837 - 0.0730991 I  \\
     & 2 &  	 				        &  			     	     & 0.234855 - 0.1229610 I 
\\ \colrule
\hline
     & 0 &  0.135881 - 0.0352487 I  & 0.228421 - 0.0350311 I   & 0.320544 - 0.0349743 I \\
-0.4 & 1 &  	  				  	& 0.222865 - 0.1062450 I   & 0.316498 - 0.105515 I  \\
     & 2 &  			   			   &  			     	      & 0.308788 - 0.177778 I
\\ \colrule
\hline
     & 0 &   0.173271 - 0.0481029 I  & 0.289911 - 0.0477632 I  & 0.406319 - 0.0476741 I \\
-0.3 & 1 &  				  		 & 0.281872 - 0.1450910 I  & 0.400437 - 0.1439510 I \\
     & 2 &  				   			&  			     	      & 0.389321 - 0.2429190 I
\\ \colrule
\hline
     & 0 &   0.214218 - 0.0629859 I & 0.356772 - 0.0624889 I  & 0.499397 - 0.0623572 I \\
-0.2 & 1 &   				   		& 0.345728 - 0.1901120 I  & 0.491277 - 0.1884430 I  \\
     & 2 &    				   	   &  			             & 0.476060 - 0.3184740 I
\\ \colrule
\hline
     & 0 &  0.258657 - 0.0799073 I  & 0.428832 - 0.0792154 I  & 0.599513 - 0.079030 I \\
-0.1 & 1 &      			  		& 0.414250 - 0.2413500 I  & 0.588739 - 0.239019 I  \\
     & 2 &  				  		   &  			             & 0.568719 - 0.404524 I
\\
\botrule
             & {\bf{0}} &  {\bf{0.306551 - 0.0988743 I}}   & {\bf{0.505966 - 0.0979492 I}} & {\bf{0.706469 - 0.0976977 I}} \\
  {\bf{0.0}} & {\bf{1}} &      			  		         & {\bf{0.487306 - 0.2988410 I}} & {\bf{0.692615 - 0.2957070 I}}  \\
             & {\bf{2}} &  				  		          &  			                  & {\bf{0.667091 - 0.5011410 I}}
\\ \colrule
\hline
    & 0 &  0.357881 - 0.119892 I   & 0.588087 - 0.118696 I  & 0.820117 - 0.118365 I \\
0.1 & 1 &      			  		   & 0.564807 - 0.362617 I  & 0.802751 - 0.358530 I  \\
    & 2 &  				  		   &  			            & 0.771026 - 0.608384 I
\\ \colrule
\hline
    & 0 &   0.412644 - 0.142963 I   & 0.675133 - 0.141459 I  & 0.940349 - 0.141034 I \\
0.2 & 1 &   				   		& 0.646694 - 0.432702 I  & 0.919034 - 0.427505 I  \\
    & 2 &    				   		&  			             & 0.880425 - 0.726304 I
\\ \colrule
\hline
    & 0 &   0.470850 - 0.168088 I   & 0.767067 - 0.166242 I  & 1.067090 - 0.165709 I \\
0.3 & 1 &  				  		    & 0.732935 - 0.509113 I  & 1.041390 - 0.502645 I  \\
    & 2 &  				   			&  			     		 & 0.995229 - 0.854935 I
\\ \colrule
\hline
    & 0 &  0.532513 - 0.195268 I     & 0.863869 - 0.193046 I  & 1.20029 - 0.192388 I \\
0.4 & 1 &  	  				  	  	 & 0.823523 - 0.591860 I  & 1.16977 - 0.583959 I  \\
    & 2 &  			   			  	 &  			     	  & 1.11541 - 0.994300 I
\\ \colrule
\hline
    & 0 &   0.597671 - 0.224495 I   & 0.965538 - 0.221869 I & 1.33994 - 0.221070 I \\
0.5 & 1 &   				  	    & 0.918468 - 0.680945 I & 1.30416 - 0.671445 I  \\
    & 2 &  	 				   		&  			     		& 1.24098 - 1.144400 I
\\ \colrule
\hline
    & 0 &     0.66636 - 0.255765 I  & 1.07209 - 0.2527090 I  & 1.48601 - 0.251753 I \\
0.6 & 1 &   				   		& 1.0178 - 0.77636100 I  & 1.44457 - 0.765098 I  \\
    & 2 &  	 				   		&  			     		 & 1.37197 - 1.305230 I
\\ \botrule
    & 0 &   0.738623 - 0.289068 I   & 1.18354 - 0.285559 I  & 1.63854 - 0.284430 I \\
0.7 & 1 &   				   		& 1.12156 - 0.878081 I  & 1.59101 - 0.864899 I  \\
    & 2 &  	 				   		&  			     		& 1.50844 - 1.476750 I
\\ \botrule
    & 0 &   0.814518 - 0.324389 I   & 1.29994 - 0.320411 I  & 1.79756 - 0.319092 I \\
0.8 & 1 &   				   		& 1.22981 - 0.986085 I  & 1.74354 - 0.970822 I  \\
    & 2 &  	 				   		&  			     		& 1.65047 - 1.658910 I
\\ \botrule
    & 0 &   0.894148 - 0.361698 I   & 1.42135 - 0.357252 I  & 1.96311 - 0.355727 I \\
0.9 & 1 &   				   		& 1.34265 - 1.100300 I  & 1.90222 - 1.082830 I  \\
    & 2 &  	 				   		&  			     		& 1.79815 - 1.851630 I
\\ \botrule
\end{tabular} 
\label{table:Third set}
}
\end{table}


\begin{table}[ph!]
\centering
\caption{
Dirac Quasinormal frequencies (varying $\xi$, $n$ and $Q_{\text{YM}}$) with $M=2$ and $Q=1$ for the  model considered in this work. 
%
}
{
\begin{tabular}{c|c|cccc} 
\toprule
$Q_{\text{YM}}$ & $n$ &  $\xi=6$ & $\xi=7$ & $\xi=8$ & $\xi=9$ 
\\ \colrule
-0.750 & 0 & \ 0.0729629\, -0.00301249 i \ & 0.0850142\, -0.00296907 i \ & 0.0971603\, -0.00295424 i \ & 0.109321\, -0.00296387 i \\
-0.750 & 1 & \ 0.0729444\, -0.00992241 i \ & 0.0842962\, -0.00857149 i \ & 0.0966158\, -0.00855755 i \ & 0.108916\, -0.00866922 i \\
-0.750 & 2 & \ 0.0705580\, -0.02257350 i \ & 0.0804725\, -0.01301790 i \ & 0.0939818\, -0.01300480 i \ & 0.107270\, -0.01352370 i \\
-0.750 & 3 & \ 0.0624926\, -0.05829030 i \ & 0.0624464\, -0.01591990 i \ & 0.0826927\, -0.01498340 i \ & 0.101112\, -0.01648440 i
\\ \botrule
-0.500 & 0 & \ 0.209378\, -0.0134652 i \ & 0.2444340\, -0.0134522 i \ & 0.279471\, -0.0134433 i \ & 0.312648\, -0.0117884 i \\
-0.500 & 1 & \ 0.208133\, -0.0408892 i \ & 0.2433130\, -0.0407294 i \ & 0.278461\, -0.0406203 i \ & 0.310857\, -0.0332341 i \\
-0.500 & 2 & \ 0.206482\, -0.0694793 i \ & 0.2416080\, -0.0689556 i \ & 0.276800\, -0.0685705 i \ & 0.300527\, -0.0453110 i \\
-0.500 & 3 & \ 0.205820\, -0.0990231 i \ & 0.2402850\, -0.0982075 i \ & 0.275169\, -0.0974804 i \ & 0.249681\, -0.0294970 i
\\ \botrule
-0.375 & 0 & \ 0.291891\, -0.0198044 i \ & 0.3431550\, -0.0209802 i \ & 0.392384\, -0.0209589 i \ & 0.441591\, -0.0209434 i \\
-0.375 & 1 & \ 0.279377\, -0.0636859 i \ & 0.3410490\, -0.0637438 i \ & 0.390459\, -0.0635008 i \ & 0.439831\, -0.0633279 i \\
-0.375 & 2 & \ 0.245147\, -0.1411890 i \ & 0.3383440\, -0.1084870 i \ & 0.387590\, -0.1076830 i \ & 0.436981\, -0.1070710 i \\
-0.375 & 3 & \ 0.232768\, -0.3048920 i \ & 0.3382160\, -0.1548990 i \ & 0.385929\, -0.1536890 i \ & 0.434534\, -0.1525470 i
\\ \botrule
-0.250 & 0 & \ 0.388768\, -0.0304489 i \ & 0.4506210\, -0.0286713 i \ & 0.515815\, -0.0285691 i \ & 0.580820\, -0.0285126 i \\
-0.250 & 1 & \ 0.385984\, -0.0926911 i \ & 0.4341380\, -0.0918719 i \ & 0.503145\, -0.0889103 i \ & 0.570812\, -0.0873669 i \\
-0.250 & 2 & \ 0.382683\, -0.1577680 i \ & 0.3931320\, -0.1980080 i \ & 0.465517\, -0.1747560 i \ & 0.539826\, -0.1615380 i \\
-0.250 & 3 & \ 0.381219\, -0.2245880 i \ & 0.3889570\, -0.4101020 i \ & 0.415487\, -0.3509280 i \ & 0.478669\, -0.2982640 i
\\ \botrule
-0.125 & 0 & \ 0.492891\, -0.0417651 i \ & 0.5761850\, -0.0420339 i \ & 0.658946\, -0.0419877 i \ & 0.741662\, -0.0419538 i \\
-0.125 & 1 & \ 0.489214\, -0.1272550 i \ & 0.5729050\, -0.1276190 i \ & 0.655925\, -0.1271720 i \ & 0.738882\, -0.1268430 i \\
-0.125 & 2 & \ 0.485093\, -0.2166200 i \ & 0.5686170\, -0.2164790 i \ & 0.651550\, -0.2151750 i \ & 0.734566\, -0.2141500 i \\
-0.125 & 3 & \ 0.482644\, -0.3081080 i \ & 0.5651730\, -0.3077690 i \ & 0.647623\, -0.3057790 i \ & 0.730300\, -0.3040200 i
\\ \botrule
0.000 & 0 &  \ 0.603087\, -0.0487277 i \ & 0.7038460\, -0.0487253 i \ & 0.804576\, -0.0487237 i \ & 0.905287\, -0.0487226 i \\
0.000 & 1 &  \ 0.598930\, -0.1466360 i \ & 0.7002720\, -0.1465090 i \ & 0.801443\, -0.1464270 i \ & 0.902499\, -0.1463700 i \\
0.000 & 2 &  \ 0.590834\, -0.2458840 i \ & 0.6932630\, -0.2452830 i \ & 0.795270\, -0.2448900 i \ & 0.896988\, -0.2446190 i \\
0.000 & 3 &  \ 0.579237\, -0.3472830 i \ & 0.6830940\, -0.3456640 i \ & 0.786242\, -0.3445970 i \ & 0.888884\, -0.3438580 i 
\\ \botrule
0.125 & 0 &  \ 0.718714\, -0.0623143 i \ & 0.8400660\, -0.0626667 i \ & 0.967206\, -0.0666488 i \ & 1.088870\, -0.0666831 i \\
0.125 & 1 &  \ 0.710312\, -0.1908250 i \ & 0.8329540\, -0.1907880 i \ & 0.960222\, -0.2026210 i \ & 1.082500\, -0.2021920 i \\
0.125 & 2 &  \ 0.699220\, -0.3283490 i \ & 0.8223860\, -0.3258910 i \ & 0.950651\, -0.3451440 i \ & 1.072850\, -0.3432170 i \\
0.125 & 3 &  \ 0.692466\, -0.4739280 i \ & 0.8137850\, -0.4686010 i \ & 0.946397\, -0.4936640 i \ & 1.065730\, -0.4903010 i
\\ \botrule
0.250 & 0 &  \ 0.852997\, -0.0831217 i \ & 0.9968790\, -0.0832199 i \ & 1.140610\, -0.0832931 i \ & 1.284230\, -0.0833504 i \\
0.250 & 1 &  \ 0.844837\, -0.2540380 i \ & 0.9895980\, -0.2531850 i \ & 1.134070\, -0.2526340 i \ & 1.278320\, -0.2522620 i \\
0.250 & 2 &  \ 0.836039\, -0.4344950 i \ & 0.9801720\, -0.4310640 i \ & 1.124660\, -0.4285530 i \ & 1.269200\, -0.4266910 i \\
0.250 & 3 &  \ 0.832752\, -0.6212510 i \ & 0.9737880\, -0.6157940 i \ & 1.116690\, -0.6111170 i \ & 1.260420\, -0.6072900 i
\\ \botrule
0.375 & 0 &  \ 0.985630\, -0.0959589 i \ & 1.152230\, -0.0963718 i \ & 1.314870\, -0.0926172 i \ & 1.480690\, -0.0929845 i \\
0.375 & 1 &  \ 0.972408\, -0.2947930 i \ & 1.140760\, -0.2941700 i \ & 1.302960\, -0.2819980 i \ & 1.470200\, -0.2821800 i \\
0.375 & 2 &  \ 0.957748\, -0.5085530 i \ & 1.125450\, -0.5039680 i \ & 1.284370\, -0.4821600 i \ & 1.452880\, -0.4800560 i \\
0.375 & 3 &  \ 0.956116\, -0.7320350 i \ & 1.117200\, -0.7247310 i \ & 1.267900\, -0.6949430 i \ & 1.435240\, -0.6890760 i
\\ \botrule
0.500 & 0 &  \ 1.124700\, -0.107931 i \ & 1.312840\, -0.107005 i \ & 1.502810\, -0.107727 i \ & 1.692600\, - 0.108272 i \\
0.500 & 1 &  \ 1.105140\, -0.333655 i \ & 1.295940\, -0.328183 i \ & 1.488330\, -0.328503 i \ & 1.679960\, - 0.328919 i \\
0.500 & 2 &  \ 1.080790\, -0.580517 i \ & 1.271820\, -0.566247 i \ & 1.465990\, -0.562685 i \ & 1.659340\, - 0.560354 i \\
0.500 & 3 &  \ 1.062770\, -0.844142 i \ & 1.250650\, -0.820802 i \ & 1.444930\, -0.811924 i \ & 1.638030\, - 0.805153 i
\\ \botrule
0.750 & 0 &  \ 1.449480\, -0.164787 i \ & 1.694720\, -0.164836 i \ & 1.939620\, -0.164873 i \ & 2.184280\, -0.164903 i \\
0.750 & 1 &  \ 1.434470\, -0.504593 i \ & 1.681130\, -0.502251 i \ & 1.927310\, -0.500681 i \ & 2.173080\, -0.499582 i \\
0.750 & 2 &  \ 1.421690\, -0.863183 i \ & 1.665940\, -0.855763 i \ & 1.911360\, -0.850134 i \ & 2.157150\, -0.845853 i \\
0.750 & 3 &  \ 1.421450\, -1.231070 i \ & 1.658920\, -1.220980 i \ & 1.900410\, -1.211710 i \ & 2.143910\, -1.203840 i
\\ \botrule
\end{tabular} 
\label{table:Fifth set}
}
\end{table}

\subsection{QNMs in the eikonal limit for scalar perturbations}

The eikonal regime is obtained when $\ell \gg 1$. In this situation, the WKB approximation becomes increasingly accurate (albeit the results appear to provide remarkably precise predictions, even for small values of $\ell$), this is the reason why we can get analytical expressions for the corresponding quasinormal frequencies. 
In concrete, when $ \ell \rightarrow \infty$, the angular momentum term dominates the expression for the effective potential, so the latter takes the form
\begin{equation}
V(r) \approx \frac{f(r) \ell^2}{r^2} \equiv \ell^2 g(r) ,
\end{equation}
where we have introduced a new function $g(r) \equiv f(r)/r^2$ for simplicity. 
Now, it is required to obtain the point at which the potential takes its maximum, in this case, labeled by $r_1$.  To maintain the article self-contained, it is required to include a few details regarding the connection between (circular) null-geodesics and the eikonal approximation, to show how the point $r_1$ can be obtained.
The standard procedure used to compute the geodesics in the spacetime \eqref{lineelement} can be consulted in \cite{Chandrasekhar:1985kt} and recent works can be found in \cite{Li:2021zct,Guo:2021bcw}. 
Summarizing, we should restrict our attention to equatorial orbits, with a Lagrangian given by the form
\begin{align}
    2\mathcal{L} &= -f(r)\,\dot{t}^2 + \frac{1}{f(r)}\dot{r}^2 + r^2\dot{\phi}^2,
\end{align}
where $\phi$ is the angular coordinate. Notice that we have taken consistently the same signature $(-,+,+,+)$. 
The generalized momenta, coming from the latter Lagrangian, are
\begin{align}
p_t &= - f(r)\,\dot{t}\equiv  - E={\rm
const}\,,\label{pt2}
\\
p_{\phi} &=r^2\,\dot{\phi}\equiv
L={\rm const}\,,\label{pvarphi2}
\\
p_r &=\frac{1}{f(r)}\dot{r}\,.\label{pr2}
\end{align}
As the  Lagrangian is independent of $t$ and $\phi$, then $p_t$ and $p_{\phi}$ are two integrals of motion. Solving
(\ref{pt2})-(\ref{pvarphi2}) for $\dot{t}$ and $\dot{\phi}$, we get
\begin{align}
\dot{\phi} &= \frac{L}{r^2},\, \qquad \dot{t}=\frac{E}{f(r)}\,.\label{tdot}
\end{align}
The Hamiltonian is given by
\begin{align} 
2{\cal H} &= 2\Bigl( p_t \dot{t}+p_{\phi} \dot{\phi}+p_r 
\dot{r}-{\cal L} \Bigl) ,
\end{align}
or, equivalently
\begin{align}
2{\cal H}&= -E\dot{t} + L\dot{\phi} + \frac{1}{f(r)}\dot{r}^2=\delta_1={\rm const}\,.
\label{2Ham}
\end{align}
Notice that $\delta_1=0$ represents null geodesics and $\delta_1=1$ describes massive particles. In what follows, we will restrict to the case $\delta_1=0$, i.e., massless particles. 
So, replacing Eq.~(\ref{tdot}) in (\ref{2Ham}) and using
the definition  $\dot{r}^2 = \mathcal{V}(r)$,  we obtain
\begin{align}
\mathcal{V}(r) &= E^2 - f(r) \frac{L^2}{r^2} \,.
\label{defVr}
\end{align}
The conditions $\mathcal{V}(r) = 0$ and $\mathcal{V}'(r)=0$ for circular null geodesics lead, respectively, to:
\begin{align}
\frac{E}{L} &= \pm 
\frac{1}{r_1} \sqrt{f(r_1)}\,,
\label{LElight}
\end{align}
and
\begin{equation}
2 f(r_1) - r_1 \frac{df(r)}{dr}\Bigg|_{r_1} = 0. \label{cirgeo} 
\end{equation} 
The last equation, Eq.~\eqref{cirgeo}, is precisely required to obtain the critical value $r_1$.

\begin{figure*}[ht!]
\centering
\includegraphics[scale=0.95]{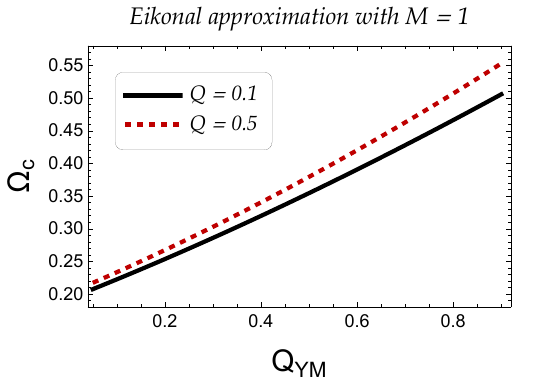} \
\includegraphics[scale=0.95]{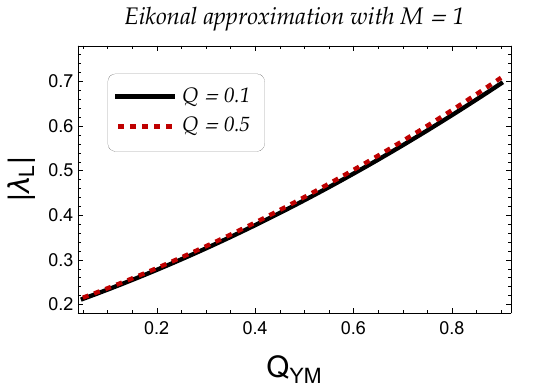} \
\caption{
QNMs in the eikonal limit:
{\bf{Left panel:}} Angular velocity vs the Yang-Mills charge assuming $Q=0.1$ and $Q=0.5$ for $M=1$.
{\bf{Right panel:}} Lyapunov exponent against the Yang-Mills charge assuming $Q=0.1$ and $Q=0.5$ for $M=1$.
}
\label{fig:eikonal} 	
\end{figure*}

The pioneering work on this topic, including the idea and formalism, can be found in Reference \cite{Cardoso:2008bp}.\footnote{For the study of QNMs in the eikonal limit beyond Einstein Relativity, we refer the reader to reference \cite{Glampedakis:2019dqh}.}
The expression for the QNMs in the eikonal regime reads:
\begin{equation} \label{omegaeikonal}
\omega(\ell \gg 1) = \Omega_c \ell - i \left(n+\frac{1}{2}\right) |\lambda_L| ,
\end{equation}
where $\lambda_L$ and $\Omega_c$ are, respectively, the Lyapunov exponent and the coordinate angular velocity at the unstable null geodesic, defined as follows \cite{Cardoso:2008bp}
\begin{align}
\lambda_L &\equiv \sqrt{\frac{1}{2} f(r_1)r_1^2 \Bigg( 
\frac{\mathrm{d^2}}{\mathrm{d}r^2} \frac{f(r)}{r^2}  
\Bigg) \Bigg|_{r = r_1} } = \pm r_1^2 \sqrt{\frac{g''(r_1) g(r_1)}{2}}, \label{Lambda}
   \\
\Omega_c &\equiv \frac{\dot{\phi}(r_1)}{\dot{t}(r_1)} = \frac{\sqrt{f(r_1)}}{r_1} = \sqrt{g(r_1)}. \label{Omega}
\end{align}
Notice that $\lambda_L$ is a measure of the rate of convergence or divergence of null rays in the ring’s vicinity, or, in other words, $\lambda_L$ is the decay rate of the unstable circular null geodesics.
In particular, for this case, we can obtain analytic exact expressions for $\{ \lambda_L, \Omega_c \}$. As they are quite involved, we show, instead, for the purpose of the present analysis approximated expressions at leading order in $Q$ and $Q_{\rm YM}$. These are given by
\begin{align}
    |\lambda_L| &\approx \frac{1}{3\sqrt{3} M} 
    \Bigg[
    \Bigg( 1 + \frac{Q^2}{18 M^2} \Bigg) 
    +
    \Bigg( 2 + \frac{Q^2}{6 M^2} \Bigg) Q_{\text{YM}}
    +
    \Bigg( 1 + \frac{Q^2}{6 M^2} \Bigg) Q^2_{\text{YM}} 
    \Bigg]
    + 
    \mathcal{O}(Q^3,Q^3_{\text{YM}}), \label{Lambda_aprox}
    \\
    \Omega_c &\approx \frac{1}{3\sqrt{3} M} 
    \Bigg[
    \Bigg( 1 + \frac{Q^2}{6 M^2} \Bigg) 
    +
    \Bigg( \frac{3}{2} + \frac{5Q^2}{12 M^2} \Bigg) Q_{\text{YM}}
    +
    \Bigg( \frac{3}{8} + \frac{5 Q^2}{ 16 M^2} \Bigg) Q^2_{\text{YM}} 
    \Bigg]
    + 
    \mathcal{O}(Q^3,Q^3_{\text{YM}}). \label{Omega_aprox}
\end{align}
Also, we can see how the $\{ \lambda_L, \Omega_c \}$ depends separately of  $\{ Q, Q_{\text{YM}} \}$. Thus, when $Q \rightarrow 0$ the corresponding functions are:
\begin{align}
   \Bigl|\lambda_L(Q \rightarrow 0) \Bigl| &=  \frac{1}{3\sqrt{3} M} 
    \Bigl( 1 + Q_{\text{YM}} \Bigl)^2,
    \\
 \Omega_c(Q \rightarrow 0) &=  \frac{1}{3\sqrt{3} M} 
    \Bigl( 1 + Q_{\text{YM}} \Bigl)^{3/2},
\end{align}
and when $Q_{\text{YM}} \rightarrow 0$ the functions are:
\begin{align}
       \Bigl|\lambda_L(Q_{\text{YM}} \rightarrow 0) \Bigl| &\approx \frac{1}{3\sqrt{3} M} \Bigg( 1 + \frac{Q^2}{18M^2}\Bigg),
       \\
        \Omega_c(Q_{\text{YM}} \rightarrow 0) & \approx \frac{1}{3\sqrt{3} M} \Bigg( 1 + \frac{Q^2}{6M^2}\Bigg).
\end{align}
The WKB approximation of 1st order produces the same expression mentioned above for 
$\{ \Omega_c, \lambda_L \}$, see for instance \cite{Ponglertsakul:2018smo}.  
Be aware that photons, in the presence of nonlinear electromagnetic sources, follow null trajectories, but 
of an effective geometry \cite{Breton:2016mqh,Breton:2017hwe,Chaverra:2016ttw}. Therefore,  the formulas for $\Omega_c$ and $\lambda_L$ remain unchanged. 
From \eqref{omegaeikonal}, we notice that the Lyapunov exponent determines the imaginary part of the modes while the angular velocity determines the real part of the modes.
In concrete, analytic expressions for the spectrum are found to be
\begin{align}
\omega_{R}(\ell \gg 1) &\equiv \text{Re}(\omega) = \Omega_c \ell\,,
\\
\omega_{I}(\ell \gg 1) &\equiv \text{Im}(\omega) = - \left(n+\frac{1}{2}\right) |\lambda_L|\,,
\end{align}
To quantify better the impact of both charges on the angular velocity and the Lyapunov exponent, given by Eqs. \eqref{Lambda_aprox}-\eqref{Omega_aprox} respectively, we vary such parameters for a fixed mass $M=1$. The result is shown in Fig.\eqref{fig:eikonal}. We can infer from the plot the following:
\begin{itemize}
    \item The angular velocity, $\Omega_c$, exhibits a monotonic increase with the Yang-Mills charge, $Q_{\text{YM}}$, for the two electric cases considered here ($Q=0.1$ and $Q=0.5$). Consequently, since $\Omega_c$ is proportional to the real part of $\omega$ ($\Omega_c \propto  \text{Re}(\omega)$), it follows that the real part of $\omega$ increases as well.
    \item The absolute value of the Lyapunov exponent, $|\lambda_L|$, shows a monotonic increase for the considered numerical values of the charges ($Q=0.1$ and $Q=0.5$) as the Yang-Mills charge is varied. Furthermore, we observe that the Lyapunov exponent remains relatively unchanged when the electric charge $Q$ is varied for small values of $Q_{\rm YM}$, resembling the behavior observed in the case of the standard Reissner-Nordstr\"{o}m BH. 
    Finally, since $|\lambda_L|$ is proportional to the negative of the imaginary part of $\omega$ ($|\lambda_L| \propto -\text{Im}(\omega)$), we can conclude that the BH is stable against scalar perturbations, given that $\text{Im}(\omega) < 0$.
\end{itemize}
The features observed in the eikonal limit, where the spectra were analytically computed, are consistent with the trends depicted in Figures \eqref{fig:1A},\eqref{fig:2},\eqref{fig:5} and \eqref{fig:6}, where the frequencies were numerically computed for low angular degrees $\ell=1,2,3$.
Thus, we can conclude that the behavior is quite similar to the results computed using the WKB approach in the previous section.

\section{Black hole shadow}

\begin{figure*}[ht!]
\centering
\includegraphics[scale=0.55]{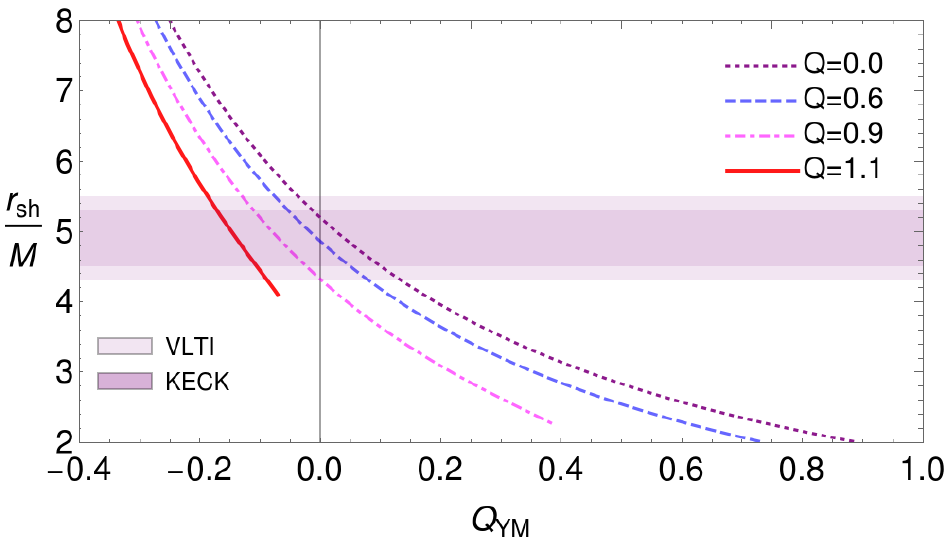} 
\includegraphics[scale=0.7]{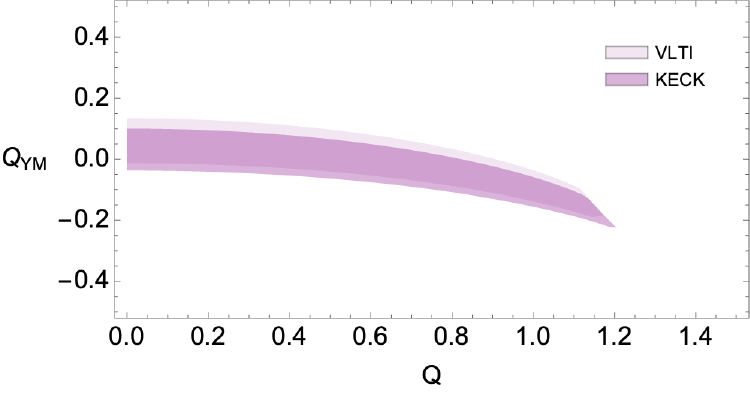} 
\caption{
\textbf{Left panel:} shadow radius for the EMPYM solution Eq.~(\ref{eqnshadow}) as a function of the Yang-Mills charge $Q_{\rm YM}$ for different values of the electric charge $Q$, as indicated in the legend. Dotted (purple) curve corresponds to the purely power Yang-Mills case, i.e. with vanishing electric charge $Q=0$, as given by Eq.~(\ref{eqnshadowPYM}). A maximum value of $Q=1.1$, solid (red) curve, requires a very small negative value of $Q_{\rm YM}$. The interception with the vertical solid line represents the shadow radius of the standard RN BH Eq.~(\ref{eqnshadowRN}). 
\textbf{Right panel:} Allowed parameter space, taking into considering the constraints given by Eqs.~(\ref{eqnKECK}) and (\ref{eqnVLTI}), for KECK and VLTI, respectively.
}
\label{fig:shadow} 	
\end{figure*}

The BH shadow is a dark region surrounded by circular orbits of photons known as \textit{photon sphere}. The radius of the photon orbit is defined as  
\begin{equation}
r_{\rm ph} = 2g_{tt}(r_{\rm ph})\left( \frac{dg_{tt}}{dr} \right)^{-1} \bigg\rvert_{r=r_{\rm ph}}\label{rphoton}.
\end{equation}
Considering the metric solution Eq.~(\ref{metricfunction}), we can compute the photon radius. It reads
\begin{equation}
 r_{\rm ph} =  \frac{3 M + \sqrt{9 M^2 - 8 Q^2 (1 + Q_{\rm YM})}}{2 (1 + Q_{\rm YM})}.\label{eqnphoton}
\end{equation}
The radius of the BH shadow is defined as the minimal impact parameter of photons escaping from the BH \cite{Perlick:2021aok}. Photons with smaller impact parameters will eventually cross the horizon and fall onto the singularity. The shadow radius can be calculated in terms of the photon sphere as \cite{Psaltis:2007rv}
\begin{equation}
r_{\rm sh} = \frac{r_{\rm ph}}{\sqrt{-g_{tt}(r_{\rm ph})}}.\label{rshadow}
\end{equation}
Considering again the metric solution Eq.~(\ref{metricfunction}) and Eq.~(\ref{eqnphoton}), the shadow radius for this class of modified RN BH is 
\begin{equation}
  r_{\rm sh} = \frac{\sqrt{2} M  Q \left(\sqrt{9 - 8 Q^{2} (1 + Q_{\rm YM})}+3\right)}{(1 + Q_{\rm YM}) 
  \sqrt{ 
  4 Q^{2} (1 + Q_{\rm YM}) + \sqrt{9 - 8 Q^{2} (1 + Q_{\rm YM})}-3 }}.
  \label{eqnshadow}
\end{equation}
It is illustrative to see some limit cases. For instance, when $Q_{\rm {YM}}\to 0$, we recover the standard RN BH solution
\begin{equation}
    r_{\rm sh} = \frac{\sqrt{2} M Q \left(\sqrt{9-8 Q^2}+3\right)}{\sqrt{4 Q^2+\sqrt{9-8 Q^2}-3 }},\label{eqnshadowRN}
\end{equation}
while the limit $Q\to 0$ leads to the purely power Yang-Mills case
\begin{equation}
    r_{\rm sh} =\frac {3 \sqrt{3} M}{1 + Q_{\rm YM}}.\label{eqnshadowPYM}
\end{equation}
This can also be interpreted as a modification of the shadow radius for the Schwarzschild BH solution.
The EHT collaboration has imaged the central BH at the center of the elliptical galaxy M87 \cite{EventHorizonTelescope:2019dse,EventHorizonTelescope:2019ths}. This data is consistent with theoretical predictions of GR for the shadow of the Kerr BH.  These unprecedented observations were followed by the image of the Sgr A$^\star$ \cite{EventHorizonTelescope:2022xqj,EventHorizonTelescope:2022wkp}, with a bright ring also consistent with a Kerr BH geometry. For Sgr A$^{\star}$, The constraints on the shadow size are
\begin{equation}
4.5M \lesssim r_{\rm sh} \lesssim  5.5M,\label{eqnKECK}
\end{equation}
for the Keck, and  
\begin{equation}
4.3M \lesssim r_{\rm sh} \lesssim  5.3M,\label{eqnVLTI}
\end{equation}
for VLTI telescope \cite{EventHorizonTelescope:2022xqj}. By using these observational values, we can place constraints on the ($Q,Q_{\rm YM}$) parameter space through Eq.~(\ref{eqnshadow}). This is shown in the right panel of Fig.~\ref{fig:shadow}. The existence of a negative Yang-Mill charge $Q_{\rm YM}\approx-0.17$ allows for a maximum electric charge $Q\approx1.1$ that is consistent with both VTLI and KECK data. It is worth noting that, in contrast, the maximum allowed charge for the standard RN case is $Q\approx0.9$ in agreement with \cite{EventHorizonTelescope:2021dqv}. To illustrate these findings, we present the behavior of the shadow radius (Eq.(\ref{eqnshadow})) as a function of the Yang-Mill charge for specific values of the electric charge $Q$ in the left panel of Fig~\ref{fig:shadow}. The dotted curve represents the case $Q=0$, corresponding to the purely Yang-Mills scenario. This case yields an allowed range of $-0.013 \lesssim Q_{\rm YM} \lesssim 0.134$ and $-0.037 \lesssim Q_{\rm YM} \lesssim 0.100$, which is consistent with VLTI and KECK data, respectively. These data impose stringent constraints on the considered scenario. As the electric charge $Q$ increases, the allowed range shifts towards negative values of $Q_{\rm YM}$. For instance, for the maximum value $Q\approx 1.1$, the allowed range becomes slightly wider with $-0.171 \lesssim Q_{\rm YM} \lesssim -0.087$. In general, for larger values of $Q$, $Q_{\rm YM}$ must take very small negative values to be in agreement with the observational data. Finally, notice that the intersection of the vertical solid line with all curves denote the standard RN case, except for the case $Q=0$, which corresponds to the Schwarzschild case with a shadow radius of $r_{\rm sh}/M=3\sqrt{3}\approx5.196$.

While this paper was being prepared, a similar study of the QNMs and BH shadow has been done for the Einstein power Yang-Mills case with a positive cosmological constant\cite{Gogoi:2023ffh}, considering a power $3/4<p<3/2$ and positive values of the associated charged. Consequently, direct comparison with our findings is not possible since we took $p=1/2$ and included the Maxwell term. However, we do observe a similar trend in the QNMs, as shown in all our plots, as $Q$ increases in this case, even for larger values of the power.

\section{Conclusions} \label{intr}

Testing the properties of BH provides a valuable approach to unravel the nature of gravity theories in the strong field regime. In particular, the final stage of binary BH mergers is characterized by quasi-normal modes, which depend primarily on the properties of the BH and, consequently, on the underlying gravitational theory. By studying these quasi-normal modes, we can gain insights into the fundamental nature of gravity and test the predictions of different gravitational theories in the extreme conditions near BHs.
Thus, in the present paper, we have addressed to main issues: 
i) we investigate how an Einstein-Maxwell power-Yang Mills BH responds to scalar and Dirac perturbations in four-dimensional spacetime for the interesting case of a power $p=1/2$.
ii) we analyze the behavior of the BH's shadow in terms of its electric and Yang-Mills charges, deriving observational constraints from data related to Sgr A$^*$.

In particular, we have depicted graphical representations of the effective potential barrier against the radial coordinate, $r$, varying the set of parameters $\{ Q_{\text{YM}}, \ell (\xi), Q, M \}$ individually while keeping the others fixed. 
Subsequently, we have computed the QNMs of scalar perturbations both numerically (employing the WKB semi-analytic approximation) and analytically (in the eikonal limit as  $\ell \rightarrow \infty$). 
In addition, we have computed the QNMs of Dirac perturbations numerically by using the WKB semi-analytic approximation of {\bf{6th order}}. 

We thoroughly examined the influences of the electric charge $Q$, the Yang-Mills charge $Q_{\text{YM}}$, the overtone number $n$, and the angular degree $\ell$ for scalar perturbations (or $\xi$ for Dirac perturbations). Our results reveal the following:
\begin{itemize}
    \item From the QNMs computations, and for the range of parameters used, we can ensure the BH is stable against scalar and Dirac perturbations. The latter is true because $\text{Im}(\omega) < 0$.
    \item From Fig. \eqref{fig:1} we observe that $\text{Re}(\omega)$ is more sensitive to the changes when $Q_{\text{YM}}$ increases, i.e., the impact of a Yang-Mills "charge" on QNMs is only relevant for positive values of $Q_{\text{YM}}$. 
\end{itemize}
On the other hand, we have also calculated the shadow radius for this class of BHs and examined particularly the influence of the Yang-Mills charge $Q_{\rm YM}$. We found that, for a given electric charge $Q$, the shadow radius is a monotonically decreasing function of the Yang-Mills charge $Q_{\rm YM}$. Thus, large and negative values of $Q_{\rm YM}$ lead to an increase in the shadow size, while positive values significantly reduce the shadow radius. This effect can be constrained by comparing with the observed values of the shadow radius of Sgr A$^{\star}$ obtained from VLTI and KECK telescopes. As both positive and negative values are allowed within the current precision, the shadow radius can be either larger or smaller compared to the Schwarzschild case. However, as the electric charge increases, $Q_{\rm YM}$ must take smaller negative values to remain consistent with the observational data. Moreover, by satisfying the current bound on the shadow radius, we are able to impose constraints on the parameter space  ($Q,Q_{\rm YM})$, as illustrated in Fig.~\ref{fig:shadow}. Therefore, the observational data does not rule out the possibility of a BH possessing both electric and gauge charges, with the latter being of a topological nature. Further investigations of BHs, particularly in the context of gravitational wave physics, will help to more robustly test the theoretical predictions associated with this class of BHs. The impact of both the electric and magnetic charges on the image formation of this BH, using various accretions models, will be valuable in distinguishing the distinct characteristics of this BH. This idea was recently explored in the context of pure power Yang-Mills case \cite{Chakhchi:2022fls}. Thus, using current and future observational data of BHs is a promising strategy in the field of gravitational physics, enabling further insights and advancements in our understanding of BH phenomena.

\section*{ACKNOWLEDGMENTS}

The authors would like to thank the reviewers for their invaluable comments, which significantly improved the quality of this manuscript.
A.R. is funded by the Generalitat Valenciana (Prometeo excellence programme grant CIPROM/2022/13) and by
the Maria Zambrano contract ZAMBRANO 21-25 (Spain).
G. G. acknowledges financial support from Agencia Nacional de Investigaci\'on y Desarrollo (ANID), Chile, through the FONDECYT postdoctoral Grant No. 3210417.

\appendix*
\label{Apendice}

\section{lower orders for the WKB method}
We show here that the sixth-order WKB method is sufficient to provide accurate results. Although higher-order corrections are possible, they do not guarantee a better approximation in all cases. We have made a comparison with the lower orders of the WKB approximation, and the results show that the sixth order is the most appropriate for the range of parameters considered. Adopting the expression
   \begin{align}
       \Delta_k &= \frac{1}{2} \Bigl|\omega_{k+1} - \omega_{k-1}\Bigl|,
   \end{align}
   to estimate the error for $\omega_k$ \cite{Konoplya:2019hlu},
we summarize our results in table \ref{tab:my_label} to demonstrate that it is sufficient to compute the 6th order for the present background. 
\begin{table}[h!]
\centering
\begin{adjustbox}{max width=\textwidth}
\begin{tabular}{c|c|c|c|c|c|c}
$k$ & $\Delta_k (Q_{YM}=0.0)$ & $\Delta_k (Q_{YM}=0.1)$ & $\Delta_k (Q_{YM}=0.2)$ &	$\Delta_k (Q_{YM}=0.3)$ & $\Delta_k (Q_{YM}=0.4)$ & $\Delta_k (Q_{YM}=0.5)$ \\
2&8.26$\times10^{-3}$  &1.05$\times10^{-2}$ &1.30$\times10^{-2}$ &1.59$\times10^{-2}$ &1.91$\times10^{-2}$ &2.26$\times10^{-2}$ \\
3&9.85$\times10^{-4}$ &1.31$\times10^{-3}$ &1.69$\times10^{-3}$ &2.15$\times10^{-3}$ &2.68$\times10^{-3}$ &3.28$\times10^{-3}$ \\
4&8.12$\times10^{-5}$ &1.13$\times10^{-4}$ &1.53$\times10^{-4}$ &2.02$\times10^{-4}$ &2.61$\times10^{-4}$ &3.31$\times10^{-4}$ \\
5&5.76$\times10^{-6}$ &8.51$\times10^{-6}$ &1.21$\times10^{-5}$ &1.69$\times10^{-5}$ &2.28$\times10^{-5}$ &3.03$\times10^{-5}$ \\
6&1.35$\times10^{-6}$ &2.11$\times10^{-6}$ &3.17$\times10^{-6}$ &4.59$\times10^{-6}$ &6.59$\times10^{-6}$ &8.87$\times10^{-6}$
\end{tabular}
\end{adjustbox}
\caption{
Error $\Delta_k$ for the order $k$ of scalar quasinormal modes assuming $\ell=3$, $M=1$ and $Q=0.1$. We observe that $k=6$ has the smaller error in all cases.
}
\label{tab:my_label}
\end{table}

\bibliography{ref}
\bibliographystyle{apsrev}
\end{document}